\documentclass[aps,nofootinbib,showpacs,preprintnumbers,amsmath,amssymb]{revtex4}

\usepackage{epsfig}
\usepackage{amsfonts}
\usepackage{amsmath}
\usepackage{amssymb}
\usepackage{graphics}
\usepackage{graphicx}

\def\no{\nonumber}

\begin{document}

\vspace{1cm}

\preprint{USM-TH-294, arXiv:1110.2545v2}

\title{Explicit solutions for effective four- and five-loop QCD running coupling}

\author{Gorazd Cveti\v{c}}
 \email{gorazd.cvetic@usm.cl}
\affiliation{Department of Physics and 
Centro Cient\'{\i}fico-Tecnol\'ogico de Valpara\'{\i}so,
Universidad T\'ecnica Federico Santa Mar\'{\i}a, Casilla 110-V, 
Valpara\'{\i}so, Chile}

\author{Igor Kondrashuk}
 \email{igor.kondrashuk@ubiobio.cl}
\vskip 5mm  
\affiliation{Departamento de Ciencias B\'asicas,  
Universidad del B\'\i o-B\'\i o, 
Campus Fernando May, Casilla 447, Chill\'an, Chile}

\begin{abstract}
We start with the explicit solution,
in terms of the Lambert $W$ function, 
of the renormalization group equation (RGE)
for the gauge coupling in the supersymmetric 
Yang-Mills theory described by the well-known NSVZ $\beta$-function.
We then construct a class of $\beta$-functions for which the RGE can be
solved in terms of the Lambert $W$ function. 
These $\beta$-functions are expressed in
terms of a function which is a truncated Laurent series 
in the inverse $u$ of the gauge coupling $a \equiv \alpha/\pi$. 
The parameters in the Laurent series can be 
adjusted so that the first coefficients
of the Taylor expansion of the $\beta$-function 
in the gauge coupling $a$ reproduce
the four-loop or five-loop QCD (or SQCD) $\beta$-function.
\vskip 1cm
\noindent Keywords: Exact RGE solution, 
Lambert function, Cardano equation
\end{abstract}

\maketitle   

\section{Introduction}
\label{sec:intr}

QCD running coupling $a(Q^2)$ in QCD is defined as 
$a(Q^2) \equiv \alpha_S(Q^2)/\pi \equiv g^2(Q^2)/4\pi^2$ 
where  $\alpha_S(Q^2)$ is the strong coupling parameter. 
It is the  solution of the renormalization group equation  (RGE)
\begin{eqnarray}
\frac{da}{d \log Q^2} = \beta(a) \equiv 
- \beta_0 a^2 - \beta_1 a^3  - \beta_2 a^4 - \cdots \ ,
\label{beta} 
\end{eqnarray}
in which the coefficients  $\beta_0$ and $\beta_1$ are
universal ($\beta_0=(11 - 2 n_f/3)/4$, $\beta_1=(102-38 n_f/3)/16$), and
$\beta_j$ ($j \geq 2$) characterize the chosen renormalization scheme (RSch). 
In this paper we use the notation $t = \log (Q^2/\Lambda^2)$ and $c_j = 
\beta_j/\beta_0$ and the previous equation becomes 
\begin{eqnarray}
\frac{da}{d t} = \beta(a) \equiv - \beta_0 a^2 (1  + c_1 a  + c_2 a^2 + c_3 a^3 + \cdots) \ .
\label{beta1}
\end{eqnarray}
The analytic structure of the running coupling in the complex $Q^2$ plane
that corresponds to the $\beta$-function of the type     
\begin{eqnarray}
\beta(a) &=& - \beta_0 a^2 \frac{1  + (c_1 - c_2/c_1) a }{1 - (c_2/c_1)a}
\label{beta3lGardi}
\\
& = & - \beta_0 a^2 \left( 1 + c_1 a + c_2 a^2 + (c_2^2/c_1) a^3 + (c_2^3/c_1^2) a^4 +
\cdots \right)
\label{b3lGarexp}
\end{eqnarray}
has been investigated thoroughly in ref.~\cite{Gardi:1998qr}. The solution has 
been reduced to Lambert function W, that allowed the authors of 
ref.~\cite{Gardi:1998qr} to study the cuts of analytic continuation in the 
complex plane at the effective three-loop level. 
By choosing the value of the coefficient $c_2$ in the $\beta$-function (\ref{beta3lGardi})
accordingly, this $\beta$-function can agree with any chosen $\beta$-function up to the
three-loop level, as seen from the expansion eq.~(\ref{b3lGarexp}).
Thus, up to the three-loop level, the exact solution of 
ref.~\cite{Gardi:1998qr}\footnote{
The exact two-loop solution, i.e., for $\beta(a) = - \beta_0 a^2 (1 + c_1 a)$, 
in terms of the Lambert $W$ function, was analyzed and used in 
refs.~\cite{Gardi:1998qr,Magradze:1998ng}.}
can be used as the running coupling $a(Q^2)$. The latter depends, in addition, 
on the given initial condition or, equivalently, on the QCD scale $\Lambda$).

In this paper we extend the effective three-loop solution of ref.~\cite{Gardi:1998qr} 
for the running coupling to the effective four-loop and five-loop solutions. 
We further show that the scale $\Lambda$ always enters in the result as 
an argument of the Lambert function. 
In section \ref{sec:NSVZ} we present the exact solution to the RGE of the 
NSVZ $\beta$-function \cite{Novikov:1983uc}, in terms of the Lambert $W$ function.
In section \ref{sec:ans} we propose an ansatz for a class of $\beta$-functions 
that generalizes the NSVZ $\beta$-function \cite{Novikov:1983uc}.
In sections \ref{sec:4l} and \ref{sec:5l} we present solutions to the RGE's of
this class of $\beta$-functions, for the cases when
the Taylor expansion reproduces the (arbitrarily chosen) coefficients of the
four- and five-loop $\beta$-functions, respectively.
We show that the RGE's of such class of $\beta$-functions have solutions 
in terms of the Lambert $W$ function. Further, we show that
solving the RGE's of this class of the effective four- and five-loop
$\beta$-functions reduces to finding the roots of a quadratic or cubic equation, 
respectively. In the case of the effective four-loop solution (for $c_3 \geq 0$),
we present in subsection \ref{subsec:num4l} detailed numerical results 
of the evaluations of our formulas, for the ${\overline {\rm MS}}$ RSch choice of
$c_2$ and $c_3$ coefficients (with $n_f=3$). In section \ref{sec:concl} we present 
the conclusions. 

\section{NSVZ $\beta$-function}
\label{sec:NSVZ}

The NSVZ $\beta$-function has been found from instanton calculus and 
for ${\cal N}=1$ supersymmetric Yang-Mills theory takes the form 
\cite{Novikov:1983uc}
\begin{eqnarray}
\beta_\alpha = 
-\frac{\alpha^2}{4\pi}\frac{3N}{1-\frac{N}{2\pi}\alpha},  \label{1}
\end{eqnarray}
where $\alpha = g^2/4\pi$ and the gauge group is $SU(N)$.
The number of colors is $N.$ This model is called supersymmetric QCD (SQCD). 
This result does not 
suppose that there are any flavors in the theory, 
that is the theory includes gluons and their superpartners 
gluinos. At present, this is the only nontrivial $\beta$-function known in all 
the number of loops. 

Initially, the $\beta$-function (\ref{1}) has been found in the second entry
of Ref.~\cite{Novikov:1983uc}.
The construction in that paper has been based on the fact that in 
supersymmetric Yang-Mills theory
the axial anomaly, the anomaly of the trace of the energy-momentum tensor and 
the supersymmetry anomaly form the components of
a supermultiplet. The divergence of the axial current is one of two terms 
of a component of the 
supermultiplet. Another term of the same component is proportional to the 
right-hand side of the relation 
for the axial anomaly. The coefficient of proportionality between the 
divergence of the axial current 
and the component of the supermultiplet can be interpreted as a 
$\beta$-function and coincides with 
NSVZ  $\beta$-function (\ref{1}) found later in 
Ref.~\cite{Novikov:1983uc}, first entry.

In our notation ($a \equiv \alpha/\pi$) the NSVZ $\beta$-function takes the form: 
\begin{eqnarray}
\frac{d a}{d t} = \beta(a) \equiv -\frac{a^2}{4}\frac{3N}{1-\frac{N}{2}a}.  \label{NSVZ}
\label{NSVZ2}
\end{eqnarray}
Eq. (\ref{NSVZ2}) is an ordinary differential equation of the first order 
and can be solved in terms of the Lambert ${\rm W}$ function. 
Since we will need, in the next sections, to use the procedure of deriving  
the solution, we write the procedure in detail. 

\begin{figure}[htb] 
\begin{minipage}[b]{.49\linewidth}
\centering\includegraphics[width=70mm]{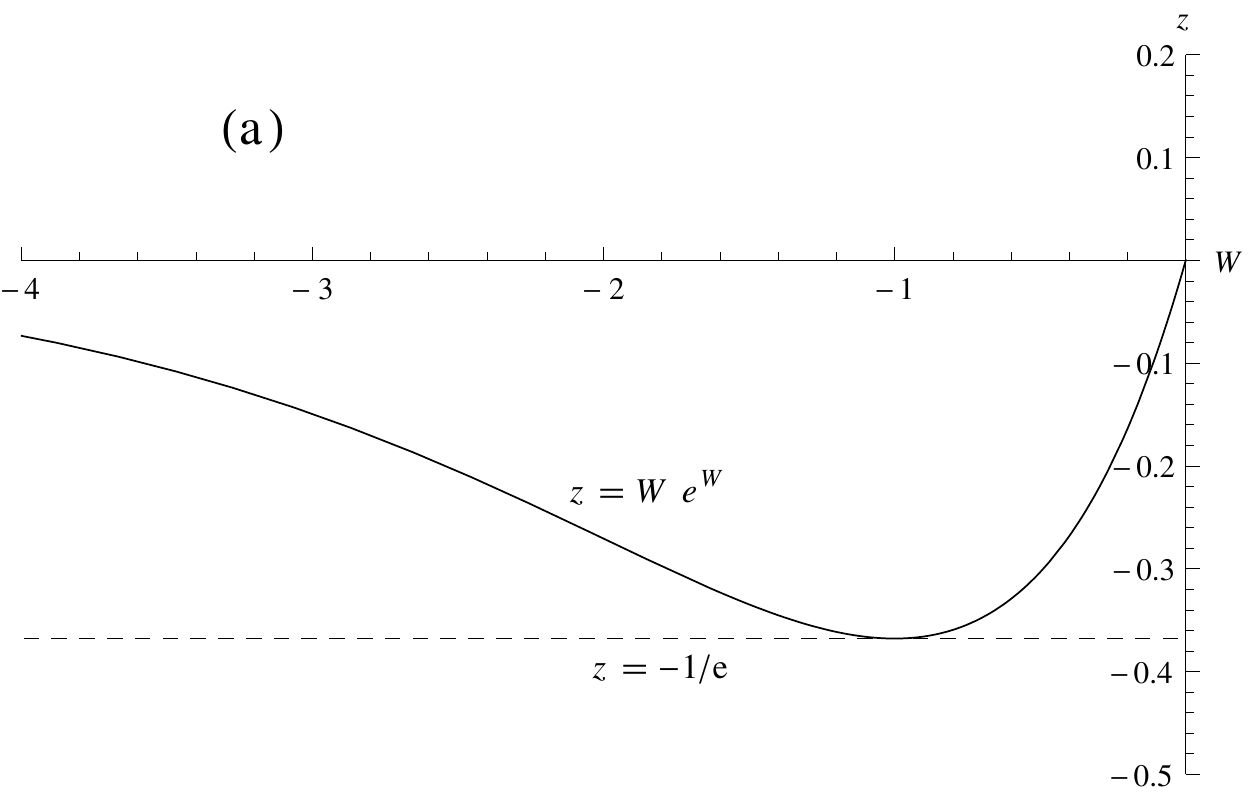}
\end{minipage}
\begin{minipage}[b]{.49\linewidth}
\centering\includegraphics[width=70mm]{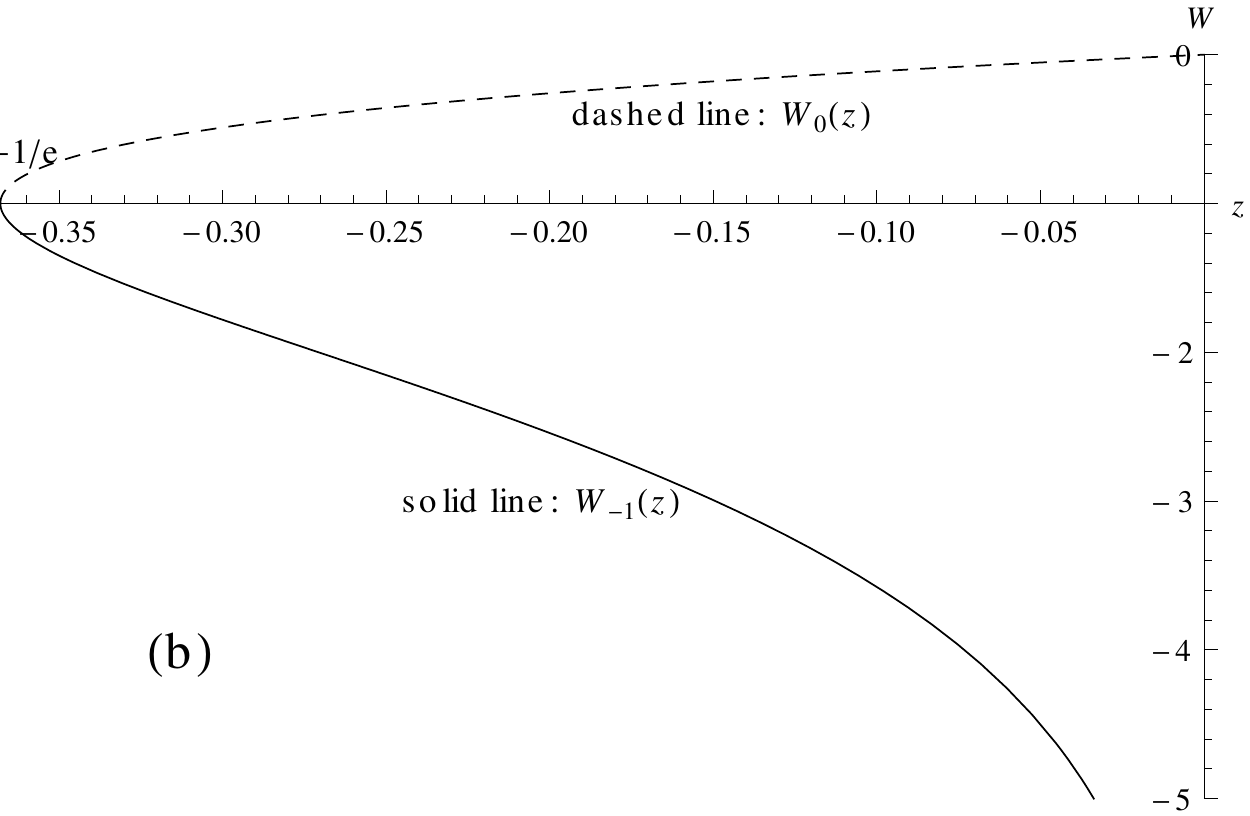}
\end{minipage}
\vspace{-0.2cm}
 \caption{\footnotesize  (a) The inverse Lambert relation $z=W e^W$, for negative $W$ (where
$z$ is negative, i.e., $Q^2$ positive); (b) The Lambert function $W=W(z)$ at negative
$z$ ($-1/e < z < 0$) - two branches are presented: $W_{-1}(z)$ and $W_0(z)$. We see that for $z \to -0$
(i.e., $Q^2 \to + \infty$), $W_{-1}(z) \to - \infty$ and $W_0(z) \to 0$.}
\label{invLL}
 \end{figure}

The chain of transformations is
\begin{eqnarray*}
\frac{d a}{dt} = a' = -\frac{a^2}{4}\frac{3N}{1-\frac{N}{2} a} \Rightarrow
-\frac{a'}{a^2} = \frac{1}{4}\frac{3N}{1-\frac{N}{2} a} \Rightarrow\\
\left(\frac{1}{a}\right)' = \frac{1}{4}\frac{3N}{1-\frac{N}{2} a} \Rightarrow
\left(\frac{2}{N a}\right)' = \frac{3/2}{1-\frac{N a}{2}} \Rightarrow
u' = \frac{3/2}{1-1/u},
\end{eqnarray*}
where the substitution $u(t) = 2/(N a(t))$ is done. The solution of the last equation 
can be found in the following way:
\begin{eqnarray*}
u' = \frac{3/2}{1-1/u} \Rightarrow u'(1-1/u) = 3/2 \Rightarrow (u-\ln u)' = 3/2 \\
\Rightarrow u-\ln u = \frac{3}{2}t + C \Rightarrow e^u/u = e^{\frac{3}{2}t+C} \Rightarrow -\frac{1}{We^W} =
 e^{\frac{3}{2}t+C},
\end{eqnarray*}
where $W = - u$ and $C$ is a constant. As a result we obtain 
\begin{eqnarray}
We^W = - e^{-\frac{3}{2}t-C} \equiv z.  \label{z}
\end{eqnarray}
Equation $We^W = z$  defines the Lambert $W$ function for the inverse relation $W = W(z).$ 
Taking into account
all the previous steps, the solution to the NSVZ equation is 
\begin{eqnarray}
u(t) &=& - W\left(-e^{-\frac{3}{2}t-C}\right) 
\nonumber\\
\Rightarrow a(t) = \frac{2}{N u(t)} 
&=& -\frac{2}{N}  \frac{1}{W\left(-e^{-\frac{3}{2}t-C}\right)}.  \label{2}
\end{eqnarray}
To interpret this result physically, one needs to analyze the inverse relation 
$W = W(z)$ of the relation $We^W = z(W).$  
The function $z(W)$ has a minimum at the point 
$W=-1$ equal to $-1/e.$ The derivative 
$z'(W)$ at this point is zero (see Fig.~\ref{invLL}(a))
and this is the branching point for the 
analytic continuation of the inverse function $W(z)$ to the complex plane. 
Let's consider first only the real values of the argument $z$ of  
function $W(z)$. In order to construct the inverse function $W(z)$,  we should 
choose between the part of ${\mathbb R}$ to the left of the minimum and the part of 
${\mathbb R}$ to the right of the minimum, 
since we need to have one-to-one correspondence between the argument and  
the function for any given interval. 

To make this choice, we impose the physical requirements, such as positivity, 
reality, continuity, and the asymptotic freedom for the running 
coupling parameter $a(t)$ in eq. (\ref{2}). The interval of 
$W \in ~ ]-\infty,-1]$ fits these criteria, because $W$ is negative in that 
interval ($a(Q^2)$ is positive).
Moreover, when  $W \rightarrow -\infty$, we have 
$z \rightarrow 0^-,$ $a(Q^2) \rightarrow 0^+$, and this represents 
the asymptotic freedom of the coupling constant $a(Q^2)$. 

The interval to the right of the minimum, $W \in ~ [-1,\infty [$, 
does not fit most of these criteria, 
since the coupling constant becomes discontinuous 
(infinite) at the point $W=0,z=0$.

\begin{figure}[htb] 
\begin{minipage}[b]{.49\linewidth}
\centering\includegraphics[width=70mm]{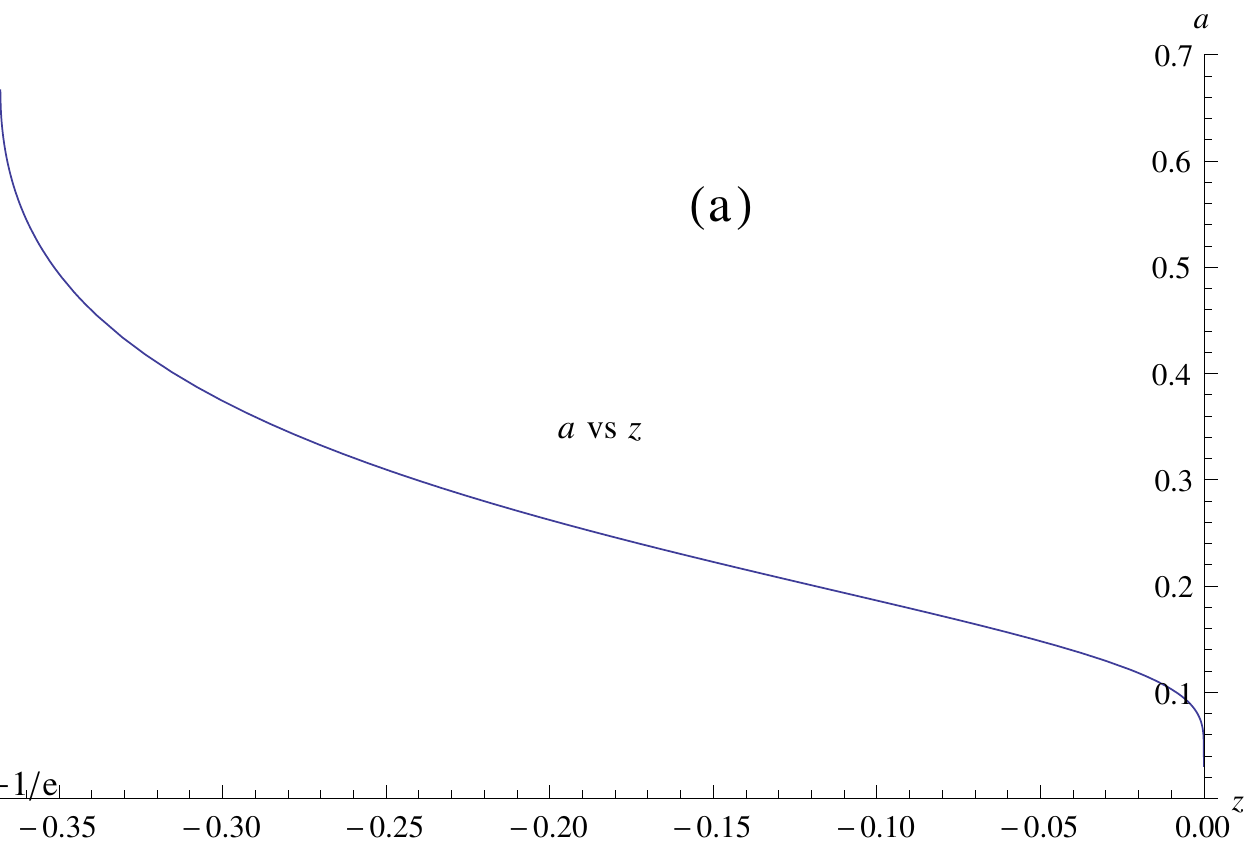}
\end{minipage}
\begin{minipage}[b]{.49\linewidth}
\centering\includegraphics[width=70mm]{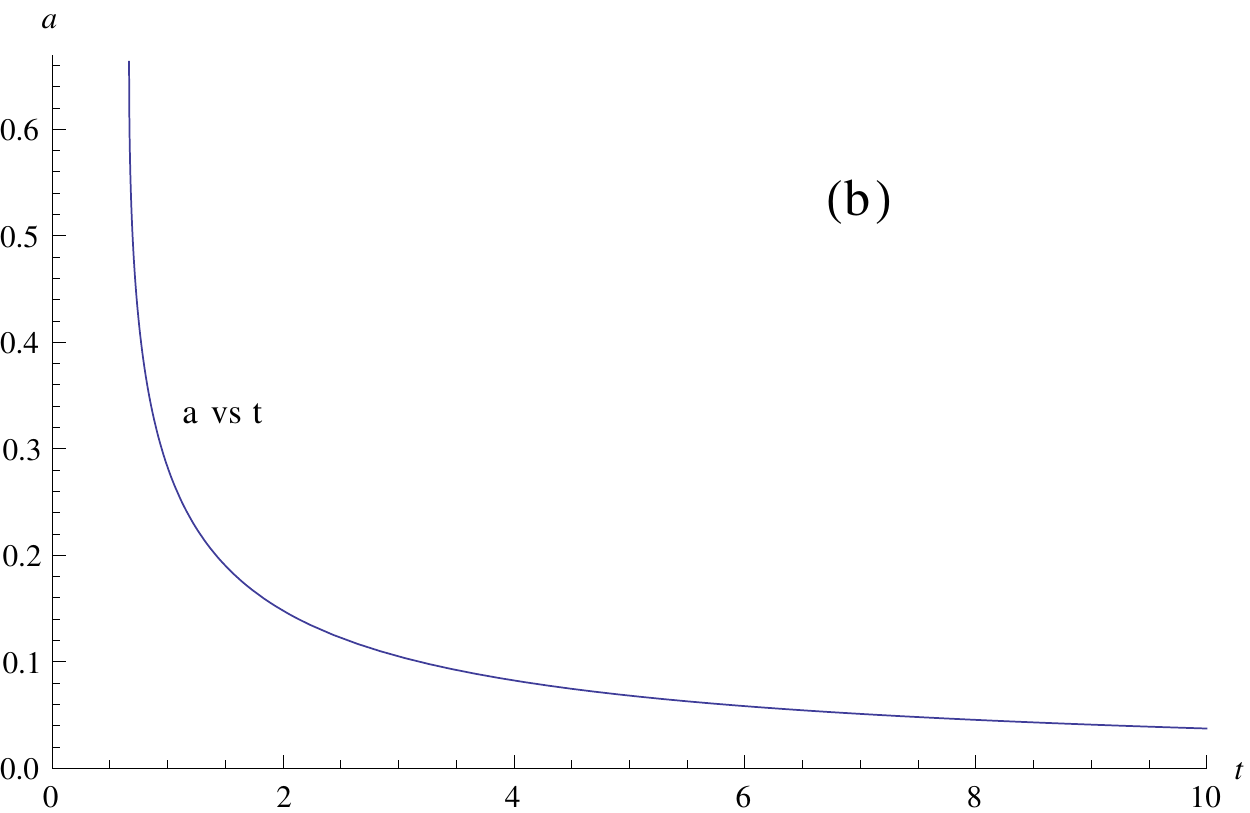}
\end{minipage}
 \caption{\footnotesize  (a) Coupling $a$ of Eq.~(\ref{2}), with $W_{-1}$
(and $N=3$), as a function of
negative $z$ ($-1/e < z < 0$); (b) $a$ as a function of $t = -(2/3) \ln(-z) =
\ln(Q^2/\Lambda^2)$, for $2/3 < t < \infty$, corresponding to
$-1/e < z < 0$. The scale $\Lambda^2$ is such that $C=0$, 
cf.~also Eq.~(\ref{z2}). At $t=2/3$ (i.e., $Q^2=\Lambda^2 e^{2/3}$) 
a Landau singularity appears, the value of $a$ there is $2/N$ ($=2/3$).}
\label{avszt}
 \end{figure}
Thus, we have to choose the interval of $W \in ~ ]-\infty,-1]$, i.e., the
branch $W_{-1}$ of the Lambert function - see also Figs.~\ref{invLL}
and \ref{avszt}.  
The value of $z$ in this interval is 
changing between $z(-1) = -1/e$ and $z(-\infty) = 0^-.$  
The variable $z$ defined by eq. (\ref{z}) is related 
to the momentum transfer $Q^2$ as 
\begin{eqnarray}
z = - e^{-\frac{3}{2}t-C}  = - \left(\frac{Q^2}{\Lambda^2}\right)^{-3/2} = 
- \left(\frac{\Lambda^2}{Q^2}\right)^{3/2} \ ,
\label{z2} 
\end{eqnarray}
where $\Lambda$ is an arbitrary scale. Arbitrariness of this scale 
follows from the  definition of $t = \ln (Q^2/M^2)$, where $M^2$ 
is the Pauli-Villars regulator in the work of ref.~\cite{Novikov:1983uc}.
An arbitrary constant $C$ of eq. (\ref{z2}) is typical for differential 
equation and it has been absorbed in $M^2$ to create the
scale $\Lambda^2$ in eq.~(\ref{z2}).
In QCD, the scale $\Lambda$ is identified with 
a specific physical scale, but this is not the case here
since this model does not have any physical scale at which 
the coupling parameter $a(t)$ is measured. Equation (\ref{z2}) is an explicit 
relation between $z$ and $Q^2,$ namely $z = -1/e$ corresponds to 
$Q^2 = \Lambda^2e^{2/3}$ and $z=0^-$ corresponds to $Q^2 \rightarrow \infty.$

The physical solution, eq.~(\ref{2}), can be written as 
\begin{eqnarray}
a(Q^2) = -\frac{2}{N} 
\frac{1}{W_{\mp 1}\left(-(\Lambda^2/Q^2)^{3/2}\right)}
\ , 
\label{NA}
\end{eqnarray}
where the branch $W_{-1}$ is taken for all $Q^2$ with ${\rm Im} Q^2 \geq 0$,
and $W_{+1}$ when ${\rm Im} Q^2 < 0$.\footnote{
Since $W_{+n}(z^*) = W_{-n}(z)^*$, we have $a(Q^{2 *}) = a(Q^2)^*$ as it should be.
For an analysis of the branches (partitions) $W_n$ of the Lambert
function, we refer to subsection \ref{subsec:cut}.}
Analytic structure of this function in the complex $Q^2$ plane 
has a cut for $Q^2 \in ~ ]-\infty,Q_b^2]$ where $Q_b^2 \equiv \Lambda^2 \exp(2/3)$ ($> 0$)
is the branching point. The cut of $a(Q^2)$ on the positive $Q^2$ semiaxis is
not physical, in the sense that it does not respect the analytic
properties of the spacelike observables ${\cal D}(a(Q^2))$. The latter
properties are dictated by the locality and causality of 
quantum field theories \cite{BS,Oehme}.\footnote{
In analytic QCD models, the cut part of perturbative $a(Q^2)$ on the positive 
$Q^2$ semiaxis is removed \cite{Shirkov:1997wi}, 
and the cut part on the negative $Q^2$ semiaxis
(at $|Q^2|  \stackrel{<}{\sim} \Lambda^2$) is expected to be modified in general.
For reviews of analytic QCD models, see refs.~\cite{anQCDrev} 
and references therein.}. The NSVZ $\beta$-function leads to an $a(Q^2)$,
Eq.~(\ref{NA}), with a cut on the positive axis
$Q^2 \in ~ ]0,Q_b^2]$.  At the branching point $Q_b^2 \equiv \Lambda^2 \exp(2/3)$ it is
finite
\begin{eqnarray}
a(Q_b^2 \equiv \Lambda^2 e^{2/3}) = \frac{2}{N},  \label{MNA}
\end{eqnarray}
but the derivatives of this function are singular at the branching point. 
The maximum value of the 
coupling is obtained at the branching point $Q^2=Q_b^2$, and it goes 
down monotonically when $Q^2$ increases above $Q_b^2$. 
In the limit of large number of colors the maximum value 
of the coupling parameter goes to zero  and the theory becomes a 
theory without interaction. 
If $\Lambda$ is considered to be an arbitrary unfixed scale, 
the branching point can be taken as close to $ Q^2 = 0$ as we wish.
On the other hand, if we fix $a(Q^2)$ to a specific ``initial''
value at a specific value of $Q_{\rm in}^2$, then $\Lambda$ is fixed as well.

Below the scale $Q_b^2 = \Lambda^2 \exp(2/3)$, the NSVZ model
cannot be applied, 
in the  sense that the effective vertices (proper multipoint Green function) 
and effective propagators (full two-point Green function)  
do not exist below that scale since 
the running coupling  ``does not exist'' 
in that region. 
A physical explanation for this phenomenon could be that the 
scattering processes in this model do not take place
in that region of the low momentum transfer $Q^2 < Q_b^2$.
Any experimental confirmation of the found behavior is impossible since 
the model does not contain the physical particles of the Standard Model. 
In QCD the situation is different 
since the scattering takes place at any momentum transfer 
$Q^2$ (even very small). Theoretically, the corresponding positive part 
of the cut of $a(Q^2)$ in the complex $Q^2$-plane in perturbative QCD 
appears as an artifact, because an approximate (truncated series) 
$\beta$-function is used there; and the mentioned cut is removed by 
the analytization procedure.

\section{Ansatz}  
\label{sec:ans}

The result in the previous section is not new (however, may have never
been presented in such a form).\footnote{Actually, it seemed to us improbable 
that nobody tried to solve the RGE for NSVZ $\beta$-function (\ref{1}). 
We searched through the numerous papers citing Gardi {\it et al.\/}
paper \cite{Gardi:1998qr} and did not find anything similar to 
what we finally wrote in Section \ref{sec:NSVZ} of the present paper. 
After the publication of the first version of the present paper
in arXiv, Tim Jones notified us \cite{TJones} that he solved Eq.~(\ref{1}) 
in 1983 and that he related the solution to 
the Lambert function after he saw Ref.~\cite{Gardi:1998qr}.} 
As we have mentioned in the Introduction, 
a similar type of the  $\beta$-function has been analyzed 
in ref.~\cite{Gardi:1998qr} in the context of investigation of the
analyticity properties of the running coupling in the complex $Q^2$ 
plane. The explicit solution to that RGE has been found there.\footnote{
Note that the effective three-loop $\beta$-function, Eq.~(\ref{beta3lGardi}),
reduces to the NSVZ $\beta$-function, Eq.~(\ref{NSVZ2}), when: $c_2=c_1^2$,
$c_1=N/2$, $\beta_0=3 N/4$.} 
However, in the previous section we did not 
follow step-by-step the derivation of ref.~\cite{Gardi:1998qr}. 
Now we construct an ansatz for solving the  
RGE based on a simple generalization of the procedure described in the 
previous section.  

As the first step, we modify eq.~(\ref{2}) 
\begin{eqnarray}
f(u(t)) = - W\left(-e^{-At-C}\right)  
\label{fW}
\end{eqnarray}
where $f(u)$ is an arbitrary continuous function of $u.$  Then, 
\begin{eqnarray*}
(f(u)-\ln f(u))' = A \Rightarrow  u'\frac{df(u)}{du}(1 - 1/f(u)) = A  \\
 u' = \frac{A/f_u'(u)}{1 - 1/f(u)}.
\end{eqnarray*}
The relation between $a$ and $u$ remain the same as it stands in the previous section, 
\begin{eqnarray}
a(t) = \frac{2}{B u(t)}, 
\label{au}
\end{eqnarray}
where $A$ and $B$ are some constants. Thus, we obtain  
\begin{eqnarray*}
\left(\frac{2}{B a(t)}  \right)' = 
\frac{A/f_u'(\frac{2}{B a(t)})}{1 - 1/f(\frac{2}{B a(t)})} \ ,
\end{eqnarray*}
where we recall that $t \equiv \ln(Q^2/M^2)$.
As a result, the solution to the following RGE
\begin{eqnarray}
\left(a(t) \right)' = \beta(a) = - \frac{A a^2}{2}
\frac{B/f_u'(\frac{2}{B a(t)})}{1 - 1/f(\frac{2}{B a(t)})} \ ,
\label{RGE}
\end{eqnarray}
is obtained via the formula
\begin{equation}
a(t) = \frac{2}{B} \frac{1}{F(-W(e^{-A t - C}))} \ ,
\label{at}
\end{equation}
where $F$ is the inverse of the function $f$ appearing in eq.~(\ref{fW}),
i.e., $F(f(u)) = u$.

The function $f(u)$ can be chosen arbitrarily. 
For our purpose, the ansatz will be a truncated Laurent series in $u$
with the leading term to be $u$
\begin{equation}
f(u) = u + a_0 + \frac{a_1}{u} + \cdots + \frac{a_n}{u^n} \ .
\label{fu}
\end{equation}
In this work we will consider two ans\"atze for this function,
one with $n=1$ and the other with $n=2$. The first one represents
a beta function with four real parameters ($A$, $B$, $a_0$, $a_1$),
which can be adjusted so that the expansion of $\beta(a)$ in powers of
$a$ reproduces the four-loop $\beta$-function in a given renormalization
scheme (RSch), i.e., the given coefficients $\beta_0$ and $c_j$ ($j=1,2,3$) of
the expansion (\ref{beta1}). The second one has five real parameters,
which can be adjusted to reproduce the five-loop $\beta$-function in
a given RSch. We will call the first and the second ansatz the 
``effective four-loop'' and the ``effective five-loop'' $\beta$-function,
respectively.
As we will see in the next sections, the problem of solving these RGE's 
is reduced to finding the inverse function of the function $f(u)$. This
means, in practice, finding the roots of polynomials.

\section{Effective four-loop case}
\label{sec:4l}

\subsection{Four-loop ansatz for $f(u)$}
\label{subsec:4lansf}

We take $f(u)$ in the form 
\begin{eqnarray}
 f(u) = u + a_0 + \frac{a_1}{u} \ , 
\label{AN1}
\end{eqnarray}
where $a_0,$  $a_1$ are arbitrary real numbers. We show how to 
reproduce the $\beta$-function  up to $\sim a^5$ tuning these two numbers
(and the numbers $A$ and $B$)
\begin{eqnarray*}
\beta(a) = -\beta_0 a^2~(1+c_1 a  + c_2 a^2 + c_3 a^3 ) + {\cal O}(a^6) \ .
\end{eqnarray*}
We note that $\beta_0$ and $c_1$ are universal in mass-independent
schemes, and $c_j$ ($j \geq 2$) are the parameters which characterize RSch.
Thus, we consider the $\beta$-function as determined in the previous section 
\begin{eqnarray}
\beta(a) = - \frac{A a^2}{2}\frac{B/f_u'(u)}{1 - 1/f(u)}, 
\label{betaour}
\end{eqnarray}
where $f(u)$ is given in eq.~(\ref{AN1}). We conclude that 
\begin{eqnarray}
\beta_0 =  \frac{AB}{2}. 
\label{be0}
\end{eqnarray}
With the coefficient $B$ fixed, we choose $A$ to adjust  $\beta_0$. 
Now we show how to fix the coefficient $B.$
In the fraction 
\begin{eqnarray}
\frac{1/f_u'(u)}{1 - 1/f(u)}, \label{AAfrac}
\end{eqnarray}
we change temporarily the normalization of $a$ in order 
to simplify the calculation, $u = 1/a$,
since $a$ appears only with the factor $2/B$. 
The correct normalization will be recovered afterwards,
by a simple redefinition of $a$, 
\begin{eqnarray}
u = 2/(B a) \ . 
\label{a}
\end{eqnarray}
Then we have 
\begin{eqnarray*}
f_u'(u) = 1 - \frac{a_1}{u^2}  ~~~ \Rightarrow ~~~
f_u'(1/a) = 1 - a_1 a^2,  \\
 f(u) = u + a_0 + \frac{a_1}{u}  ~~~ \Rightarrow ~~~ f(1/a) = \frac{1}{a} + a_0 + a_1 a,  \\
1/f_u'(1/a) = \frac{1}{1 - a_1 a^2 } = 1 + a_1 a^2 + {\cal O}(a^4) \ .
\end{eqnarray*}
Then, the expansion can be performed directly and we obtain 
\begin{eqnarray*}
\frac{1/f_u'(1/a)}{1 - 1/f(1/a)} = 
1 + a  + (a_1 - a_0 + 1) a^2 + (a_0 -1)^2  a^3 + {\cal O}(a^4) \ . 
\end{eqnarray*}
We now restore the original normalization for the running coupling $a$,
eq.~(\ref{a}), and require the $\beta$-function to be in a given RSch up to
four loops 
\begin{eqnarray*}
 1 + Ba/2  + (a_1 - a_0 + 1)(Ba/2)^2 + (a_0 -1)^2 (Ba/2)^3   
= 1+c_1 a  + c_2 a^2 + c_3 a^3.
\end{eqnarray*}
Thus, we deduce 
\begin{equation}
B/2 =  c_1\ , \quad (a_1 - a_0 + 1)(B/2)^2 = c_2 \ , \quad
(a_0 -1)^2  (B/2)^3 = c_3 \ . 
\label{ajcj1}
\end{equation}
The first identity determines $B=2 c_1$.
The other two identities then take the following form: 
\begin{eqnarray}
a_1 - a_0 + 1 = c_2/c_1^2 &\equiv& \omega_1 \ ,
\quad
(a_0 -1)^2 = c_3 /c_1^3  \equiv \omega_2 \ .
\label{oms}
\end{eqnarray}
The solution is
\begin{eqnarray}
a_0^{(\pm)}  &=&  \pm \sqrt{\omega_2} + 1 = \pm \sqrt{c_3/c_1^3} + 1 \ , 
\nonumber\\
a_1^{(\pm)} &=&  \omega_1 \pm \sqrt{\omega_2} = (c_2/c_1^2)\pm \sqrt{c_3/c_1^3} \ .  
\label{choice}
\end{eqnarray} 
We have to assume that $\omega_2 \geq 0$, because otherwise our ansatz
would give us a $\beta$-function with nonreal coefficients.

\subsection{The inverse function of $f(u)$}
\label{subsec:invf4l}

Knowing explicitly the coefficient $a_0,$ $a_1$ in the ansatz (\ref{AN1})
\begin{eqnarray}
-W = u + a_0 + \frac{a_1}{u}, \label{type1}
\end{eqnarray}
we rewrite it in the form 
\begin{eqnarray*}
u^2 + (a_0 + W) u + a_1 = 0,
\end{eqnarray*}
that is in its turn a usual quadratic equation. Taking into account the result for the previous subsection, the equation takes the form 
\begin{eqnarray*}
u^2 + (\pm\sqrt{\omega_2} + 1 + W) u + \omega_1 \pm\sqrt{\omega_2} = 0.
\end{eqnarray*}
The two roots for this equation are 
\begin{eqnarray}
u_{1,2} = \frac{-(\pm\sqrt{\omega_2} + 1 + W)\pm \sqrt{(\pm\sqrt{\omega_2} + 1 + W)^2 
- 4(\omega_1 \pm\sqrt{\omega_2})}  }{2}.  \label{u} 
\end{eqnarray}

\subsection{Cut structure and analyticity}
\label{subsec:cut}

We first analyze eq.~(\ref{u}) for the real value of the argument $z$ 
of the Lambert function $W(z)$.
The variable $z$ defined by eq. (\ref{z}) is related 
to the momentum transfer $Q^2$ as 
\begin{eqnarray}
z = - e^{-At-C}  = - \left(\frac{Q^2}{\Lambda^2}\right)^{-A} = 
- \left(\frac{\Lambda^2}{Q^2}\right)^{A}  = 
- \left(\frac{\Lambda^2}{Q^2}\right)^{\beta_0/c_1} \ ,
\label{zz}
\end{eqnarray}
where $\Lambda$ is an arbitrary scale. We recall that $t \equiv \ln(Q^2/M^2)$,
and the arbitrary constant $C$ has been absorbed in  $\Lambda^2$.
We will assume throughout that $\beta_0 > 0$ and $c_1 > 0$.

As we mentioned earlier, the function $z(W)$ has a minimum at 
the point $W=-1$ equal to $-1/e.$ The derivative $z'(W)$ at 
this point is zero and this is the branching point for 
the analytic continuation of the inverse function $W(z)$. 
This point remains a branching point for the function $u(t)$ 
found in eq.~(\ref{u}). 

To make the choice  between the part of ${\mathbb R}$ to the left of 
the minimum and the part of ${\mathbb R}$ to the right 
of the minimum, we impose, as in the NSVZ case of section \ref{sec:NSVZ},
the physical requirements: positivity, reality, continuity and 
asymptotic freedom for the running coupling $a(t)$ of eq.~(\ref{a}). 
The interval of $W \in ~ ]-\infty,-1]$ 
fits these criteria under some restrictions on the relation 
between the coefficients $\omega_1$ and $\omega_2$. 
When $Q^2 \to \infty$, we have $z \to 0^-$, and $W \to - \infty$; the
asymptotic freedom of $a(Q^2)$ then implies that we must have the
positive sign in front of the square root in eq.~(\ref{u})
\begin{eqnarray}
u  \left( = \frac{1}{c_1 a} \right) 
= \frac{-(\pm\sqrt{\omega_2} + 1 + W) 
+ \sqrt{(\pm\sqrt{\omega_2} + 1 + W)^2 
- 4(\omega_1 \pm\sqrt{\omega_2})}  }{2}.  
\label{u-} 
\end{eqnarray}
The interval to the right of the minimum, 
which is $W \in ~ [-1,\infty [$, does not fit these physical criteria,
for the same reasons as in the NSVZ case. In particular, when
$z \to 0^-$, this branch gives $W \to 0^-$, contravening the asymptotic freedom.

Therefore, we have to choose the interval of $W \in ~ ]-\infty,-1]$,  
with the values of $z$ between $z(-1) = -1/e$ and $z(-\infty) = 0^-$.
This means that, if $0 < \beta_0/c_1 < 2$, as in the NSVZ case of
section \ref{sec:NSVZ}, we have to choose the branch $W_{-1}$ for the
Lambert function for real $z \in (-1/e,0)$, i.e., for
$Q^2 \in (Q_b^2,+\infty)$ where $Q_b^2 \equiv \Lambda^2 \exp(c_1/\beta_0)$.
For details on this point, we refer to the later analysis of the
continuation of $a(Q^2)$ to complex $Q^2$, later in this subsection. 

The solution of the RGE (\ref{RGE}) can now be written as\footnote{
We note that if we take $a_1=0$ in eq.~(\ref{AN1}), i.e., $f(u) = u + a_0$,
we obtain $a_0 = 1 - \omega_1 = 1 - c_2/c_1^2$ and the $\beta$ function turns out
to be just the effective three-loop beta function of sec.~4 of
ref.~\cite{Gardi:1998qr} [here: eq.~(\ref{beta3lGardi})], and our
approach gives $a(Q^2) = -(1/c_1) 1/( 1 - (c_2/c_1^2) + W)$, just the same as
obtained in ref.~\cite{Gardi:1998qr}.} 
\begin{eqnarray}
a(Q^2) = -\frac{2}{c_1} \frac{1}{ (\pm\sqrt{\omega_2} + 1 + W) - \sqrt{(\pm\sqrt{\omega_2} + 1 + W)^2 
- 4(\omega_1 \pm\sqrt{\omega_2})} }. \label{NA-2}
\end{eqnarray}
Two signs are possible, coming from eq.~(\ref{choice}), 
and we will consider both options.
In both the cases the coupling $a(Q^2)$ goes down monotonically 
to zero with increasing $Q^2$.
Monotonic behavior can be checked directly from (\ref{NA-2}) 
by taking the derivative, 
or by using the relation for derivative of the 
inverse function $u'(W) = 1/W'(u).$   

Let us choose first the lower sign in eq.~(\ref{NA-2}). Provided that
\begin{eqnarray}
\omega_2  - 4(\omega_1 - \sqrt{\omega_2}) > 0 \ , 
\label{det-}
\end{eqnarray}
the branching point of $a(Q^2)$ is at $Q_b^2 \equiv \Lambda^2 \exp(c_1/\beta_0)$,
and the coupling reaches there its finite maximum
\begin{eqnarray*}
a(Q^2 = \Lambda^2 e^{c_1/\beta_0}) = \frac{2}{c_1} \frac{1}{\sqrt{\omega_2} 
+ \sqrt{\omega_2  - 4(\omega_1 - \sqrt{\omega_2})} } \ ,
\end{eqnarray*}
but the derivatives of this function are singular there. 
Thus, for the choice of the lower sign the solution to RGE (\ref{RGE}) 
that reproduces  first four coefficients  of the $\beta$-function is 
\begin{eqnarray}
a(Q^2) = \frac{2}{c_1} \Biggl[\sqrt{\omega_2} - 1 - 
W \left(- \left(\Lambda^2/Q^2\right)^{\beta_0/c_1}\right)  \Biggr. \no\\
\Biggl. + ~~\sqrt{\left(\sqrt{\omega_2} - 1 - 
W\left(- \left(\Lambda^2/Q^2\right)^{\beta_0/c_1}\right)\right)^2 
- 4(\omega_1 - \sqrt{\omega_2})} \Biggr]^{-1} \label{NNA-2}
\end{eqnarray}
where $Q^2 \in ~ ]\Lambda^2 \exp(c_1/\beta_0),\infty[.$ 

In order to extend the function (\ref{NNA-2}) to all the plane of complex
$Q^2$, we need to take into account for the
analytic structure  of the multivalued function $W(z)$ 
of complex variable $z$, which is described in 
ref.~\cite{Corless:1996zz}.  As it was done in ref.~\cite{Gardi:1998qr},
we follow ref.~\cite{Corless:1996zz} for the  division of the 
branches and also for the notation. Since  $W(z)$ is a multivalued function 
of the variable $z$, the corresponding $z$-plane has to be split in 
a multisheet Riemann surface with a cut for each sheet of the surface, 
while the complex plane of $W$ should be divided in partitions
having common borders. Each partition in $W$-plane can be 
bijectively mapped onto one of the  sheets 
of the Riemann surface of variable $z$. 
The borders of each partition transform under this map to edges of the cuts. 
The partitions (branches) are named 
$W_0,$ $W_1,$  $W_{-1},$  $W_2,$  $W_{-2},$ etc. 
The branches $W_n$ with $n < 0$ have negative imaginary parts,
and those with $n >0$ have positive imaginary parts. 
The $W_0$ is an exceptional case, it is the only 
partition that contains the positive part of the real axis of 
the $W$-plane completely. The border of the $W_0$ partition 
can be mapped to the edges of cut $z \in ]-\infty,-1/e]$, 
and the branching  point is  $z=-1/e$.
The branch choices conform to the rule of counter-clockwise 
continuity around the branching point. 
This means, for example, that the upper edge of the cut $z \in ]-\infty,-1/e]$ 
can be mapped onto the upper border of the $W_0$ partition in the $W$-plane. 
The next sheet of the Riemann surface has a double cut, 
one is the same $z \in ]-\infty,-1/e],$ and the other is $z \in ]-\infty,0]$.
The first cut corresponds to the border between 
$W_0$ and $W_{\pm 1}$ partitions, and the 
second cut $z \in ]-\infty,0]$ corresponds to the border between 
$W_{\pm 1}$ and $W_{\pm 2}$ partitions. According to the rule of the 
counter-clockwise continuity, the upper part of the cut 
$z \in ]-1/e,0]$ transforms to the border of the $W_{-1}$ partition. 
This means that $W_{0}$ and $W_{-1}$ are the only partitions that 
contain the real values of $W$. The part of the cut 
$z \in ]-1/e,0]$ corresponds to the border between the $W_{1}$ and $W_{-1}$ 
partitions. These 
two partitions have common real limit along $z \in ]-1/e,0].$  

To relate behavior of the running coupling in the complex $z$-plane and 
the complex $Q^2$-plane, the phase analysis is 
important. Here we mainly follow the lines of ref.~\cite{Gardi:1998qr}, 
and write the same notation 
$Q^2 = \vert Q^2\vert e^{i\phi}$, where $-\pi<\phi<\pi$ and $z=\vert z\vert e^{i\delta}.$ 
We consider the case $c_1 > 0$. 

The domain for the argument $z$ of the Lambert function $W(z)$ is a 
Riemann surface, it looks like a ``pie''
with many horizontal ``layers''. This is in close analogy with the 
Riemann surface for the usual logarithmic function of the complex variable. 
This analogy is not surprising since for the large values 
of $z$ the Lambert function $W(z)$ has a logarithmic asymptotic behavior. 
The partitions in the $W$-plane resemble the partitions for the complex plane 
of the logarithmic function.

Each sheet has a cut. Each cut has two edges, and one of the edges belongs 
to the sheet. The edges are mapped to the borders of the partitions 
in the $W$-plane. The edge that belongs to the sheet should be glued to 
the next upper sheet (the edge does not belong to the latter). 
The edge that does not belong to the sheet should be glued to the edge of 
the previous lower sheet (the edge belongs to the latter).  
All this is in complete analogy with the Riemann surface 
for the argument of the logarithmic function. 

The sheet of the surface with $-\pi < \delta \leq \pi$ is the domain of $z$ 
for the $W_0$ partition, while for $\pi < \delta \leq 3\pi$ we pass to the next 
domain of $z$ for the $W_1$ partition, and so on, encountering new domain
each time the phase $\delta$ of $z$ increases by $2 \pi$.
Similarly, the sheet with 
$-3\pi  <  \delta \leq -\pi$ is the domain of $z$ for the $W_{-1}$ partition, 
and so on.

As has been done in ref.~\cite{Gardi:1998qr} we consider the case 
$c_1>\beta_0/2 > 0$, which is the case valid in QCD 
(with $0 \leq n_f \leq 6$).\footnote{In ref.~\cite{Gardi:1998qr}, other cases
have been considered, too.}
As mentioned earlier, the relevant partition (branch) is $W_{-1}(z)$, and the
domain of $z = |z| e^{i \delta}$ is with the phase $-3 \pi < \delta \leq - \pi$.
For positive $\phi$ (where $Q^2 = |Q^2| e^{i \phi}$) we obtain
$\delta = -\pi-(\beta_0/c_1)\phi$, in accordance with eq.~(\ref{zz}). 
It never reaches  $W_{-1}$ partition border  at $\delta=-3\pi$, 
since $\beta_0/c_1<2$.
Similarly, for negative $\phi$ the variable $z$ is in the domain of the 
partition of $W_{1}(z)$ with the phase $\pi <  \delta \leq 3\pi$ 
and we obtain  $\delta= +\pi-(\beta_0/c_1)\phi$. In turn, it never reaches 
the border of $W_1$ partition at $\delta= 3\pi$. 
The only singularity that appears for the union of 
these partitions $W_{\mp 1}$ is the singularity at the point of the cut 
start, $z=-1/e,$ that corresponds to a singularity 
on the positive real $Q^2$ axis, at 
$Q^2=Q_b^2 \equiv \Lambda^2 \exp(c_1/\beta_0)$.

The partitions  $W_{\mp 1}(z)$ defined in the domains described above have 
the common continuous limit along the line 
$z \in ]-1/e,0]$.  On the other hand, this is not the only definition of 
the domain for the union of these partitions with the common limit along 
the line $z \in ]-1/e,0]$. In ref.~\cite{Gardi:1998qr}
the phase $\delta$ of $z$ is required to be  in the range 
$-\pi <\delta \leq \pi$. If the variable $z$ is required to be in this 
domain, then for $c_1 > \beta_0/2$ we should choose for $\phi \geq 0$
(when the partition is $W_{-1}(z)$) the phase of $z$ equal to 
$\delta = +\pi-(\beta_0/c_1)\phi$, and for $\phi<0$ (when the partition is $W_{1}(z)$) 
the phase $\delta = -\pi-(\beta_0/c_1)\phi$. In both cases it never reaches 
the borders of the partitions, which is $\delta=-\pi$ for  $W_{-1}$ 
and $\delta= \pi$ for $W_{1}$.\footnote{
In Mathematica \cite{Math8}, the functions $W_n(z)$ are called by the
command ${\rm ProductLog}[n,z]$, which are periodic in $z$ with period $2 \pi$,
because numerical evaluations always give values
$z \exp(i 2 \pi) = z$.}  
In contrast, if we had $\beta_0/2 > c_1 > 0$, 
then $\delta=\pm\pi-(\beta_0/c_1)\,\phi$ and the border of the partitions 
$W_{\pm 1}(z)$ would be reached 
at the value of the phase $\phi$ of $Q^2$: $\phi = \pm 2 (c_1/\beta_0) \pi$. 
At this point we would have to include the neighboring partitions, 
increasing the modulo of their index by one, i.e., in such a case 
it would be $W_{\pm2}.$

Analytic structure of the aforementioned function $W_{\mp 1}(z)$
in the complex $Q^2$ plane has a cut for 
$]-\infty,\Lambda^2 \exp(c_1/\beta_0)]$, which contains the entire real negative axis 
and a part of positive axis, $]0,\Lambda^2 \exp(c_1/\beta_0)]$;
the branching point for this cut is $Q_b^2 = \Lambda^2 \exp(c_1/\beta_0)$
(in analytic QCD models the part of positive axis is removed  
by analytization procedure).

The analytic structure described in the previous paragraphs is 
caused by the multivaluedness of the 
Lambert function, since the running coupling $a(Q^2)$ is a composite 
function of the Lambert function. 
However, the square root in eq.~(\ref{NNA-2}) is a multivalued function too, 
\begin{eqnarray}
\varphi(W) = \sqrt{ \left( \sqrt{\omega_2} - 1 - W\right)^2  
- 4(\omega_1 - \sqrt{\omega_2})},  \label{NNA-3}
\end{eqnarray}
and the complex $\varphi$-plane should also be divided in partitions. 
Each partition can be mapped bijectively onto the corresponding 
partition of the $W$-plane, which in its turn is bijectively mapped 
onto the entire $z$-plane that has the cut described
in the previous paragraphs corresponding to the borders of the 
$W$-partition in which we work. 
The cut in the $W$-partition, caused by the multivaluedness of 
$\varphi(W)$, starts at the point where the argument of the 
square root in $\varphi$ is zero, 
\begin{eqnarray}
\left( \sqrt{\omega_2} - 1 - W\right)^2  \leq  4(\omega_1 - \sqrt{\omega_2}) ~~~ \Rightarrow \nonumber\\
-2\sqrt{\omega_1 - \sqrt{\omega_2}} \leq ~~  \sqrt{\omega_2} - 1 - W  ~~  
\leq 2 \sqrt{\omega_1 - \sqrt{\omega_2}}, \nonumber\\
-2\sqrt{\omega_1 - \sqrt{\omega_2}} - \sqrt{\omega_2} \leq ~~   - 1 - W  ~~  
\leq 2 \sqrt{\omega_1 - \sqrt{\omega_2}} - \sqrt{\omega_2}, \nonumber\\
-1 +2\sqrt{\omega_1 - \sqrt{\omega_2}} + \sqrt{\omega_2} \geq ~~   W  ~~  
\geq -1 - 2 \sqrt{\omega_1 - \sqrt{\omega_2}} + \sqrt{\omega_2}, 
\label{5.4.FLCW}
\end{eqnarray}
and connect these two limiting real points, $W= -1 + \sqrt{\omega_2} - 
2 \sqrt{\omega_1 - \sqrt{\omega_2}}$ and $W= -1 + \sqrt{\omega_2} + 
2 \sqrt{\omega_1 - \sqrt{\omega_2}}$.  
If  $\sqrt{\omega_2} - 2 \sqrt{\omega_1 - \sqrt{\omega_2}}$ is a positive 
real value, eq.~(\ref{det-}), the cut is not in the physical 
partition of the $W$-plane corresponding to our ansatz, 
it is situated in the $W_0$ partition of the $W$-plane.
Thus, the cut structure of $a(Q^2)$, in the case of the lower
sign in eqs.~(\ref{choice}) and (\ref{NA-2}), 
is dictated by the cut of $W_{\mp 1}(z)$, and not by the cut of
$\varphi(W(z))$, i.e., it starts at $Q_b^2 = \Lambda^2 \exp(c_1/\beta_0)$.

For the choice of the upper sign in (\ref{choice}),
the result for the running coupling is 
\begin{eqnarray}
a(Q^2) = -\frac{2}{c_1} 
\frac{1}
{ (\sqrt{\omega_2} + 1 + W) - \sqrt{(\sqrt{\omega_2} + 1 + W)^2 
- 4(\omega_1 + \sqrt{\omega_2})} }. \label{NA-3}
\end{eqnarray}
In this case the analytic structure due to the Lambert function is 
the same as for the choice of the 
lower sign, but the cut due to the square root can enter 
the physical region. The maximal value of
the running coupling  in this case is reached at the left edge 
of the horizontal cut, entering the 
$W_1$-partition, produced by the square root function in the denominator. In the next subsection 
we present numerical results for evaluation of the cut structure in this case of choosing 
the upper sign.

\subsection{Numerical application of analytic formulas}
\label{subsec:num4l}

In QCD, the first two universal coefficients $\beta_0$ and $\beta_1$ 
($\equiv c_1 \beta_0$) are:  $\beta_0=(11 - 2 n_f/3)/4$ and $\beta_1=(102-38 n_f/3)/16$.
It turns out that in QCD, for all numbers of active quark flavors
($0 \leq n_f \leq 6$), we have $c_1 > \beta_0/2$, i.e., the case described
in this section (i.e., the case (c) in 
section~3 of ref.~\cite{Gardi:1998qr}).  
Therefore, our formula (\ref{NA-2}) becomes relatively simple, as
in the two-loop (c) case of ref.~\cite{Gardi:1998qr}
\begin{eqnarray}
 a^{(+)}(Q^2) & = &
\frac{2}{c_1} \left[ - \sqrt{\omega_2} - 1 - W_{\mp 1}(z) + 
\sqrt{(\sqrt{\omega_2} + 1 + W_{\mp 1}(z))^2 
- 4(\omega_1 + \sqrt{\omega_2})} \right]^{-1},
\label{ap}
\\
 a^{(-)}(Q^2) & = &
\frac{2}{c_1} \left[ \sqrt{\omega_2} - 1 - W_{\mp 1}(z) + 
\sqrt{(\sqrt{\omega_2} - 1 - W_{\mp 1}(z))^2 
- 4(\omega_1 - \sqrt{\omega_2})} \right]^{-1}. 
\label{am}
\end{eqnarray}
where $Q^2 = |Q^2| \exp(i \phi)$, and the upper indices in $W$ and $z$'s 
in eqs.~(\ref{ap})-(\ref{am}) are to be used when $0 \leq \phi < \pi$, 
the lower indices when $-\pi \leq \phi < 0$, 
and $z$ is given in eq.~(\ref{zz}). The superscripts
'$(+)$' and '$(-)$' indicate that we take the upper and the
lower sign in eqs.~(\ref{choice}) and (\ref{NA-2}), respectively.
We recall that both of these solutions (\ref{ap})-(\ref{am})
work for any choices of real $c_2$ and
nonnegative $c_3$, i.e., in a sense they are 
effective ``four-loop'' solutions of the
RGE (but with coefficients $c_4, c_5, \ldots$ depending on our choice of
$c_2$ and $c_3$). \footnote{For example,
$c_4^{(\pm)} = c_2^2 + 2 c_1 c_3 \pm  3 c_2 \sqrt{c_3 c_1}  \mp c_3 \sqrt{c_3/c_1}$.}
The corresponding formula for the (pure) two-loop case 
(i.e., with $c_2=c_3=\cdots = 0$), ref.~\cite{Gardi:1998qr}, is
\begin{equation}
a^{(2-\ell)}(Q^2) =
- \frac{1}{c_1} \frac{1}{\left[ 1 + W_{\mp 1} ( z ) \right]} \ ,
\label{a2l}
\end{equation}
and the ``three-loop'' case of the beta function 
(\ref{beta3lGardi}) is similar to the previous
(ref.~\cite{Gardi:1998qr}) 
\begin{equation}
a^{(3-\ell)}(Q^2) =
- \frac{1}{c_1} \frac{1}{\left[ 1 - (c_2/c_1^2) + 
W_{\mp 1} ( z ) \right]} \ .
\label{a3l}
\end{equation}

In order to present the numerical results of the formulas 
(\ref{ap}) and (\ref{am}),
we choose the ${\overline {\rm MS}}$ renormalization scheme with $n_f=3$ 
(low energy QCD). In this case, $\beta_0=9/4$; $c_1=16/9 = 1.77778$; 
$c_2=3863/864=4.47106$; $c_3=20.9902$. 
 
The two effective ``four-loop'' beta functions, eq.~(\ref{betaour}),
$\beta^{(\pm)}(a)$, (i.e., for the choice of 
$+\sqrt{\omega_2}$ and $-\sqrt{\omega_2}$) are
\begin{eqnarray}
\beta^{(\pm)}(a) & = & - \beta_0 a^2 \frac{\left[ 1 + (a_0^{(\pm)} c_1) a +
(a_1^{(\pm)} c_1^2) a^2 \right]}
{\left[1 + (a_0^{(\pm)}- 1) c_1 a + (a_1^{(\pm)} c_1^2) a^2 \right]
\left[ 1 - (a_1^{(\pm)} c_1^2) a^2 \right]} \ ,
\label{betapm}
\end{eqnarray}
with the constants $a_0^{(\pm)}$ and $a_1^{(\pm)}$ given by eqs.~(\ref{choice}).
As argued, expansion of both beta functions
agrees, up to terms $\sim a^5$, with the four-loop ${\overline {\rm MS}}$ beta function.
The coefficients $c_j$ for $j \geq 4$ differ, though. For example,
$c_4^{(+)} =104.43$ and $c_4^{(-)} =84.81$. 

In order to fix the scales $\Lambda$ appearing in eq.~(\ref{zz}),\footnote{
We recall that these scales will differ for each choice of the beta function:
the effective four-loop beta functions (\ref{betapm}), the effective 
three-loop beta function (\ref{beta3lGardi}), and the two-loop beta function
$\beta^{(2-\ell)}(a) = - \beta_0 a^2 (1 + c_1 a)$. Further, we note that the scales
$\Lambda$ are defined in ref.~\cite{Gardi:1998qr} in a somewhat different way
[$z = - (\Lambda^2/Q^2)^{\beta_0/c_1} (1/(c_1 e))$] than here in eq.~(\ref{zz}).}
we have to adjust the
couplings at a specific scale of reference to specific values. 
We take the approximate world average $a(M_Z^2,{\overline {\rm MS}}) \approx 0.119/\pi$,
ref.~\cite{PDG2008,PDG2010}, 
and RGE-run it (at four-loop) down to the reference scale 
$\mu^2_{\rm in} = (3 m_c)^2$ ($\approx 14.516 \ {\rm GeV}^2$). The quark threshold
matching is implemented at the three-loop level, ref.~\cite{CKS}, at
threshold scales $Q^2= 3 m_q^2$ ($q=b,c$). We thus obtain the reference value,
in ${\overline {\rm MS}}$ 
\begin{equation}
a_{\rm in} \equiv a((3 m_c)^2; n_f=3) \approx 0.07245 \ .
\label{ain}
\end{equation} 
This is the reference value we use in all our numerical 
calculations, with $n_f=3$.

It turns out that the branching point in the complex $Q^2$-plane,
where the unphysical (Landau) cut of $a(Q^2)$ starts, is somewhat lower if 
we evaluate our formulas with $+ \sqrt{\omega_2}$ than with
$- \sqrt{\omega_2}$. Therefore, we present our numerical results for
the case of  $+ \sqrt{\omega_2}$, i.e., formulas (\ref{ap}) for $a^{(+)}$ and
$\beta^{(+)}$ in eq.~(\ref{betapm}).

\begin{figure}[htb] 
\begin{minipage}[b]{.49\linewidth}
\centering\includegraphics[width=70mm]{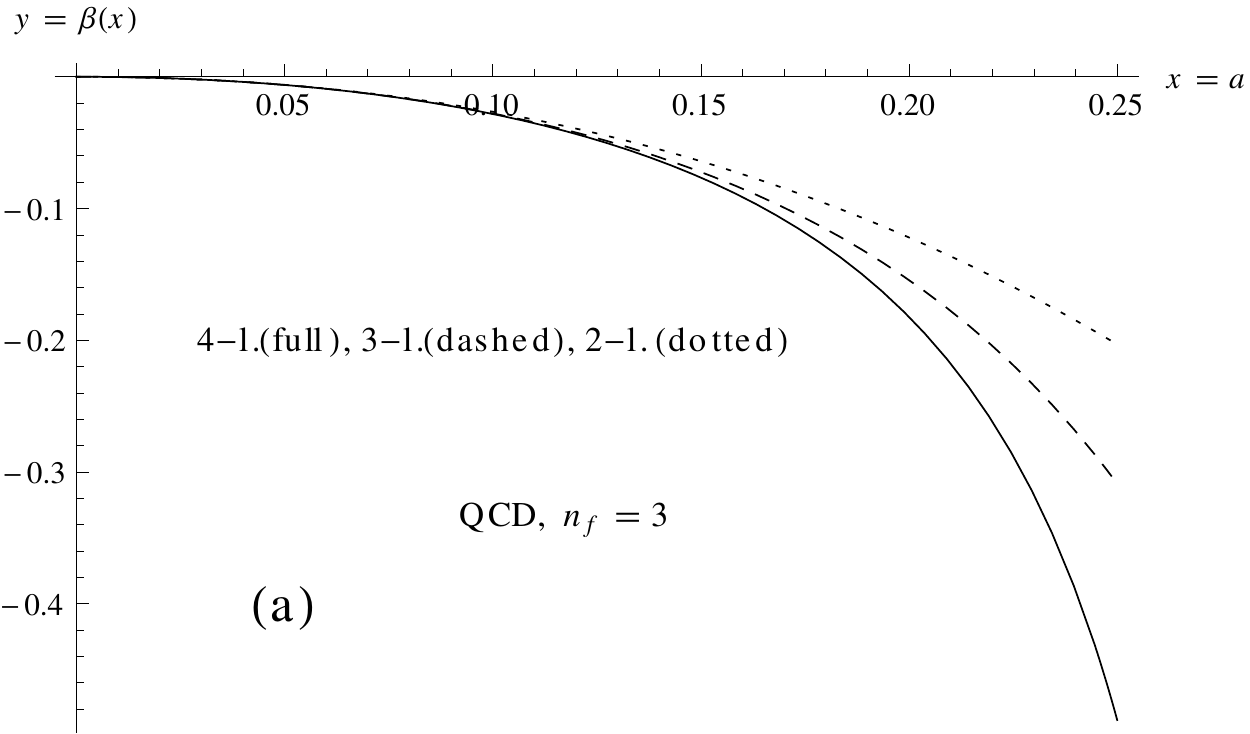}
\end{minipage}
\begin{minipage}[b]{.49\linewidth}
\centering\includegraphics[width=70mm]{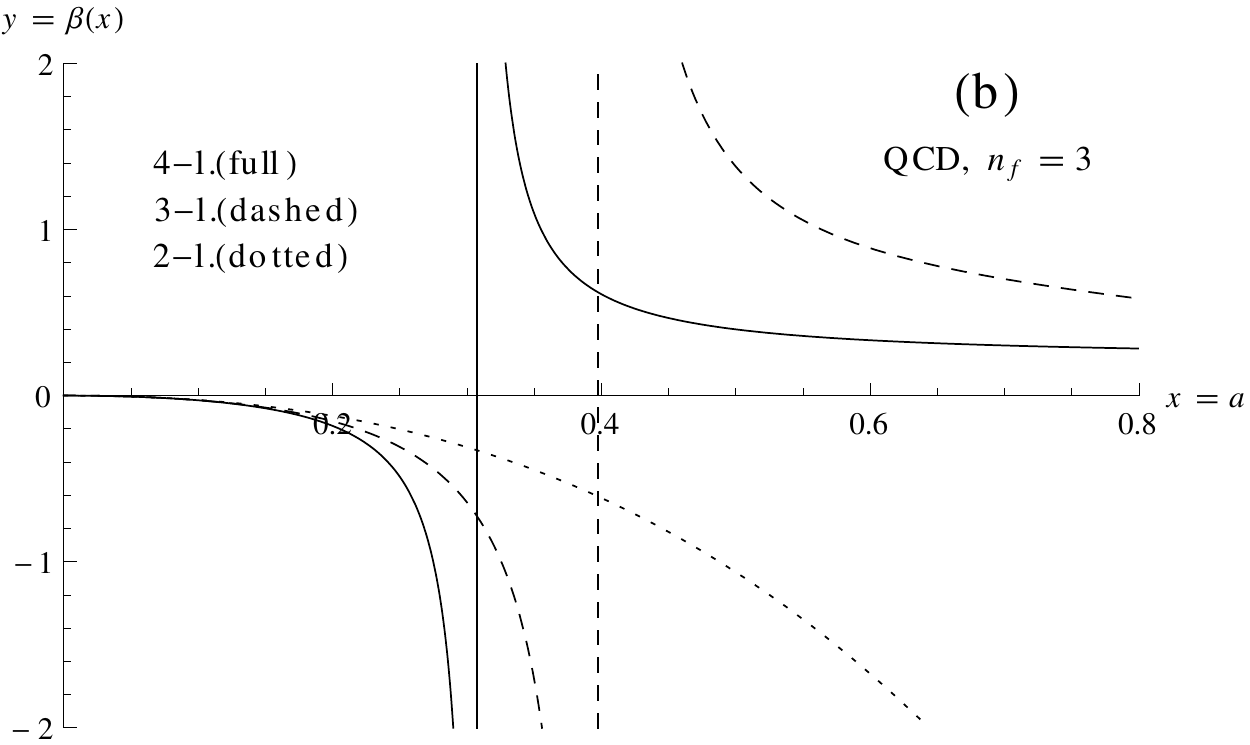}
\end{minipage}
\vspace{-0.4cm}
 \caption{\footnotesize  (a) Beta functions $\beta(a)$:
the four-loop beta function $\beta^{(+)}$ eq.~(\ref{betapm}), 
the three-loop beta function eq.~(\ref{beta3lGardi}), and the
two-loop beta function; (b) The same, for a wider scale of $a$, so
that singularities are seen.}
\label{betas}
 \end{figure}
In figs.~\ref{betas}(a),(b), we present the beta functions
$\beta(a)$, four our effective four-loop case [$\beta^{(+)}$, eq.~(\ref{betapm})], 
the effective three-loop case (\ref{beta3lGardi}), and the two-loop case.

\begin{figure}[htb] 
\begin{minipage}[b]{.49\linewidth}
\centering\includegraphics[width=75mm]{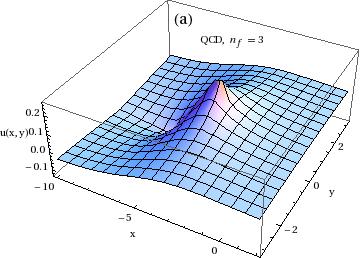}
\end{minipage}
\begin{minipage}[b]{.49\linewidth}
\centering\includegraphics[width=75mm]{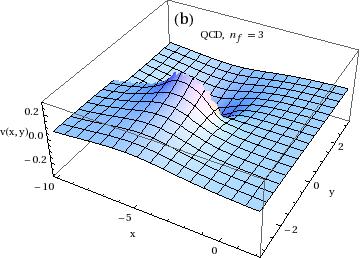}
\end{minipage}
\vspace{-0.4cm}
 \caption{\footnotesize  The (a) real and (b) imaginary part of the
effective four-loop coupling $a^{(+)}(Q^2)$, where $x = \ln(|Q^2|/\mu^2_{\rm in})$
($\mu^2_{\rm in} \approx  14.516 \ {\rm GeV}^2$) and $y={\rm Arg}(Q^2)$.}
\label{Figuv}
 \end{figure}
In figs.~\ref{Figuv}(a),(b), we present the resulting real and imaginary parts
of $a(Q^2)$ for complex $Q^2$, in the complex plane $x + i y = \ln(Q^2/\mu_{\rm in}^2)$,
i.e., where $Q^2 = \mu_{\rm in}^2 \exp(x) \exp(i y)$, with $-\pi < y < \pi$,
and $u(x,y) \equiv {\rm Re} a(Q^2)$ and $v(x,y)= {\rm Im} a(Q^2)$. The point
$x=y=0$ corresponds to $Q^2 = \mu_{\rm in}^2 \approx 14.516 \ {\rm GeV}^2$; 
the line with $y=0$ corresponds to positive $Q^2$, while the lines
with $y = \pm \pi$ correspond to negative $Q^2$.
These figures clearly indicate that there is a (Landau) singularity,
which starts at the (Landau) branching point $x_{b} \approx -3.170$, corresponding to
the branching point $Q_{b}^2 \approx 0.610 \ {\rm GeV}^2$, i.e., $Q_b \approx 0.781$ GeV.

\begin{figure}[htb] 
\centering\includegraphics[width=100mm]{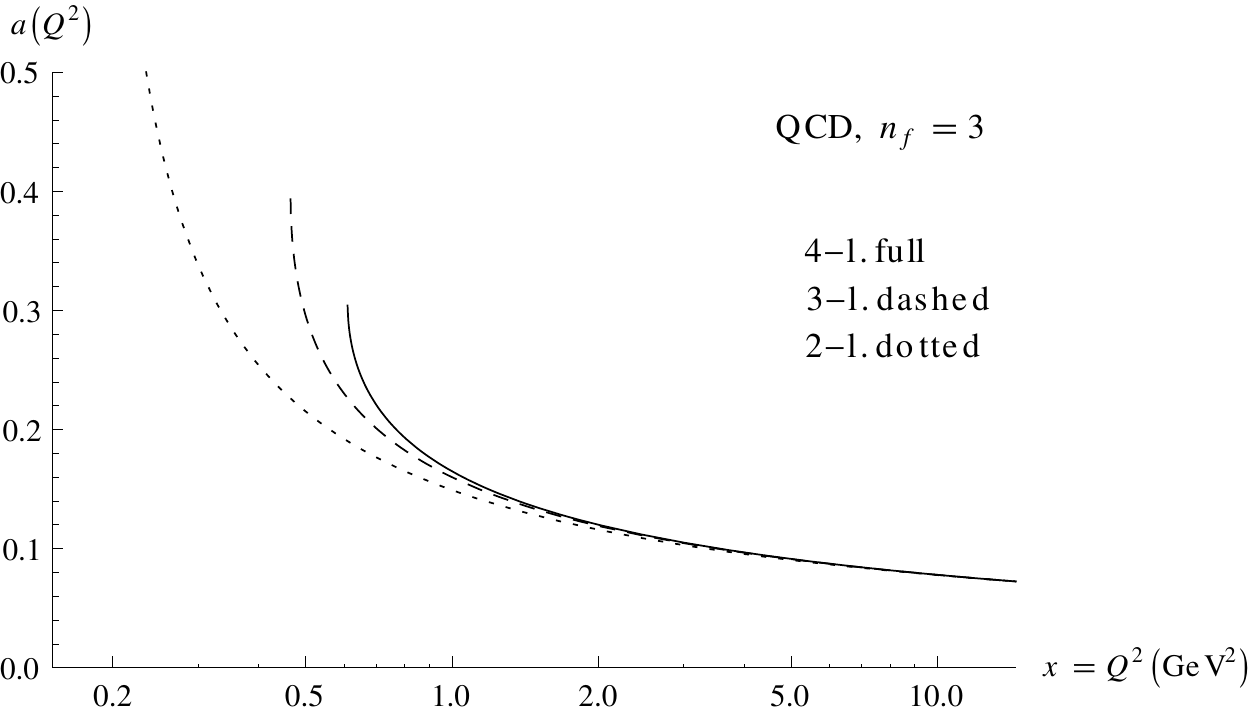}
\vspace{-0.4cm}
 \caption{\footnotesize  The effective four-loop running coupling
$a^{(+)}(Q^2)$ at positive $Q^2$. For comparison, the effective three-loop
coupling, and the two-loop coupling, are included.}
\label{aQ2}
 \end{figure}
In fig.~\ref{aQ2} we present the results for $a(Q^2)$
at positive $Q^2$ -- for our effective four-loop case (\ref{ap}), the
effective three-loop case (\ref{a3l}), and the two-loop case (\ref{a2l}).
We recall that these couplings were adjusted so that they all agree
with the value of $0.07245$
at the reference scale $\mu^2_{\rm in} = (3 m_c)^2 \approx 14.516 \ {\rm GeV}^2$.
The resulting $\Lambda$ scales appearing in the variables
$z$ of  eq.~(\ref{zz}), are: $\Lambda \approx 0.282, 0.459, 0.122$ GeV,
for the two-loop, the effective three-loop, and the effective four-loop
case, respectively.

\begin{figure}[htb] 
\begin{minipage}[b]{.49\linewidth}
\centering\includegraphics[width=70mm]{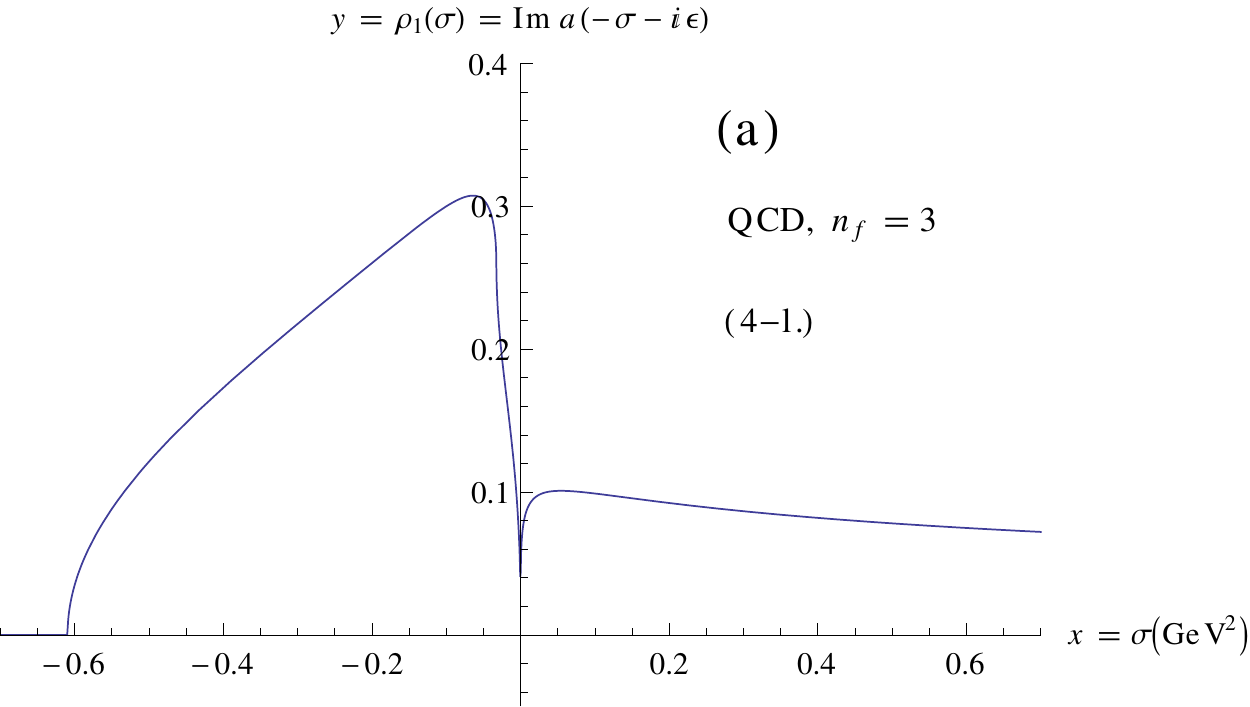}
\end{minipage}
\begin{minipage}[b]{.49\linewidth}
\centering\includegraphics[width=70mm]{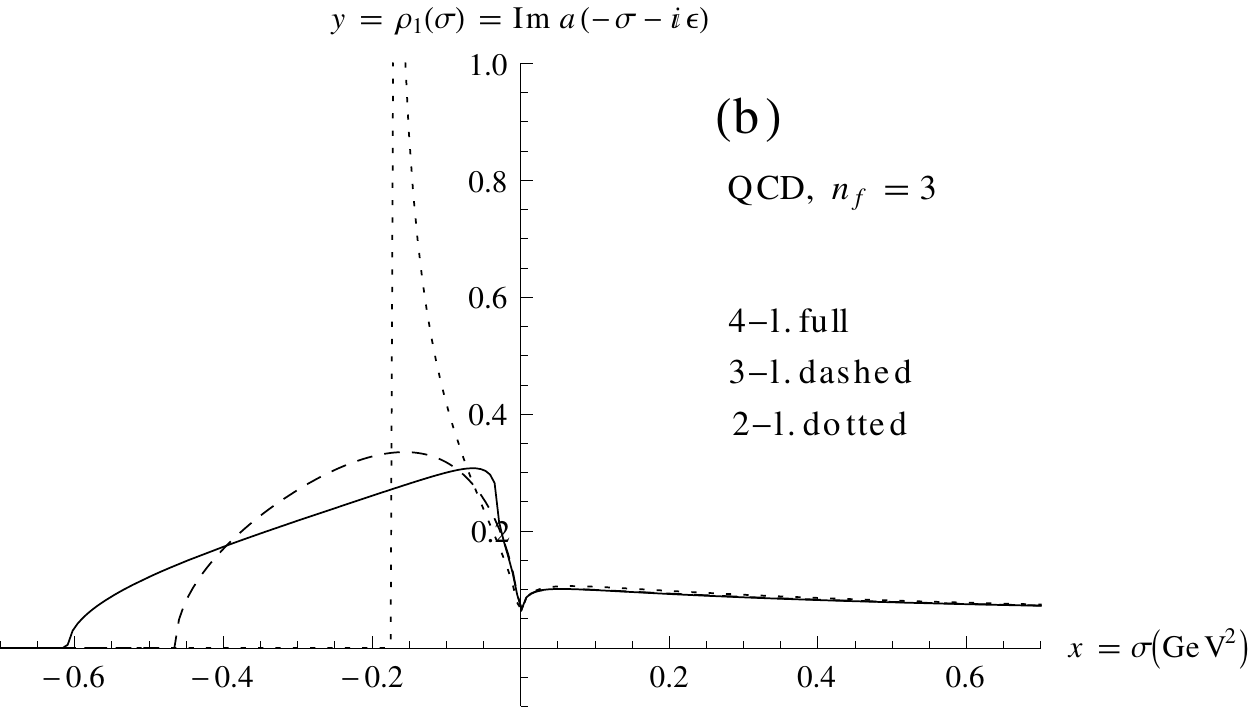}
\end{minipage}
\vspace{-0.4cm}
 \caption{\footnotesize  (a) The discontinuity function 
$\rho_1(\sigma)= {\rm Im} a^{(+)}(Q^2=-\sigma - i \epsilon)$ 
as a function of $\sigma$; (b) for comparison, $\rho_1(\sigma)$
for the effective three-loop case, and for the two-loop case, are included.}
\label{rho14l}
 \end{figure}

In fig.~\ref{rho14l}(a) the discontinuity function 
$\rho_1(\sigma) = {\rm Im} a(-\sigma - i \epsilon)$ is presented. 
The Landau branching point
$\sigma_b = - Q_b^2 \approx -0.61 \ {\rm GeV}^2$ is clearly visible. 
In fig.~\ref{rho14l}(b)
the comparison of this result with those of the effective three-loop case
and the two-loop case are presented. 
From here we can see that the
branching point $Q_b^2$ increases when the (effective) loop level increases:
$Q_b^2 \approx 0.175, 0.465, 0.610 \ {\rm GeV}^2$ ($Q_b \approx 0.418, 0.682, 0.781$ GeV)
in the two-loop, the effective three-loop, 
and the effective four-loop case, respectively. 
It is interesting that the value of the coupling at the respective
branching point is infinite in the two-loop case (also $\rho_1$ is
infinite there), and finite 
in the effective three and four-loop cases (with the values there: 
$a_b \approx 0.398, 0.307$, respectively); 
this can be seen in fig.~\ref{aQ2}.\footnote{If we use for
$\beta(a)$ the power series in $a$ truncated at the
three (four) loop-level, and the same $a_{\rm in}$ value 
as in eq.~(\ref{ain}), the numerical integration of the RGE 
gives for the branching point the value $Q_b^2 \approx 0.288 (0.394)
\ {\rm GeV}^2$; the values $a(Q_b^2)$ are infinite in such cases.}

\begin{figure}[htb] 
\begin{minipage}[b]{.49\linewidth}
\centering\includegraphics[width=70mm]{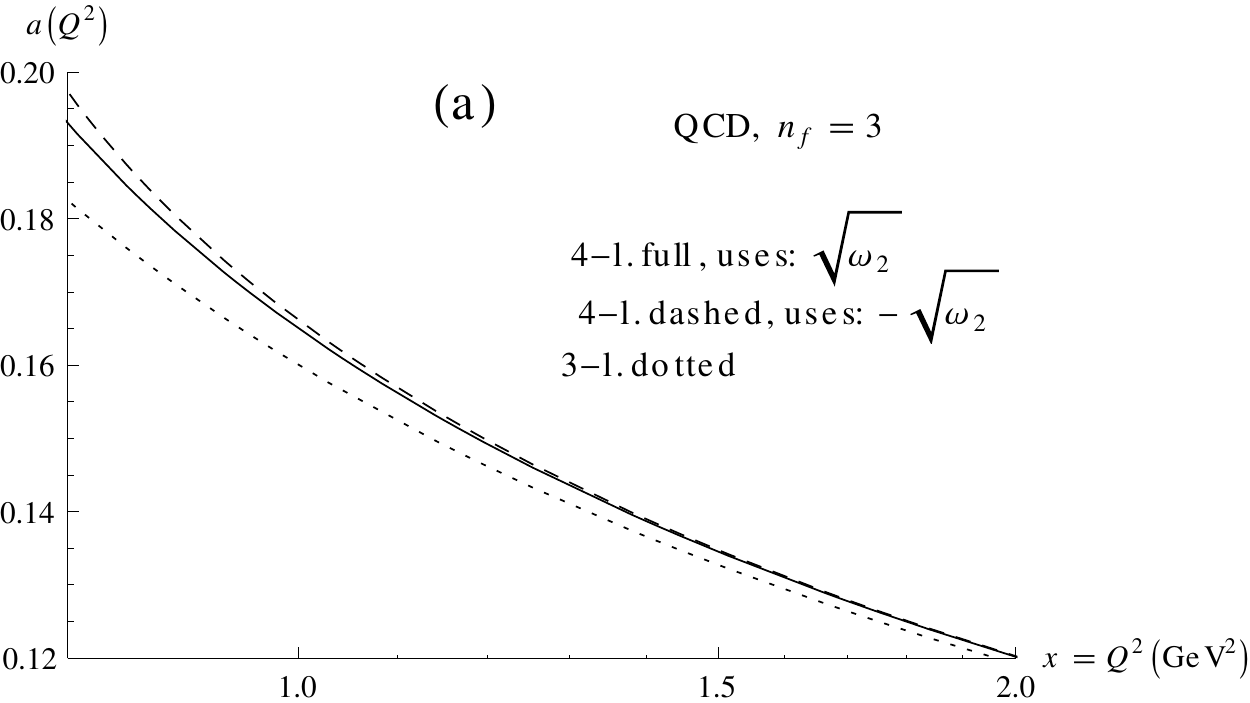}
\end{minipage}
\begin{minipage}[b]{.49\linewidth}
\centering\includegraphics[width=70mm]{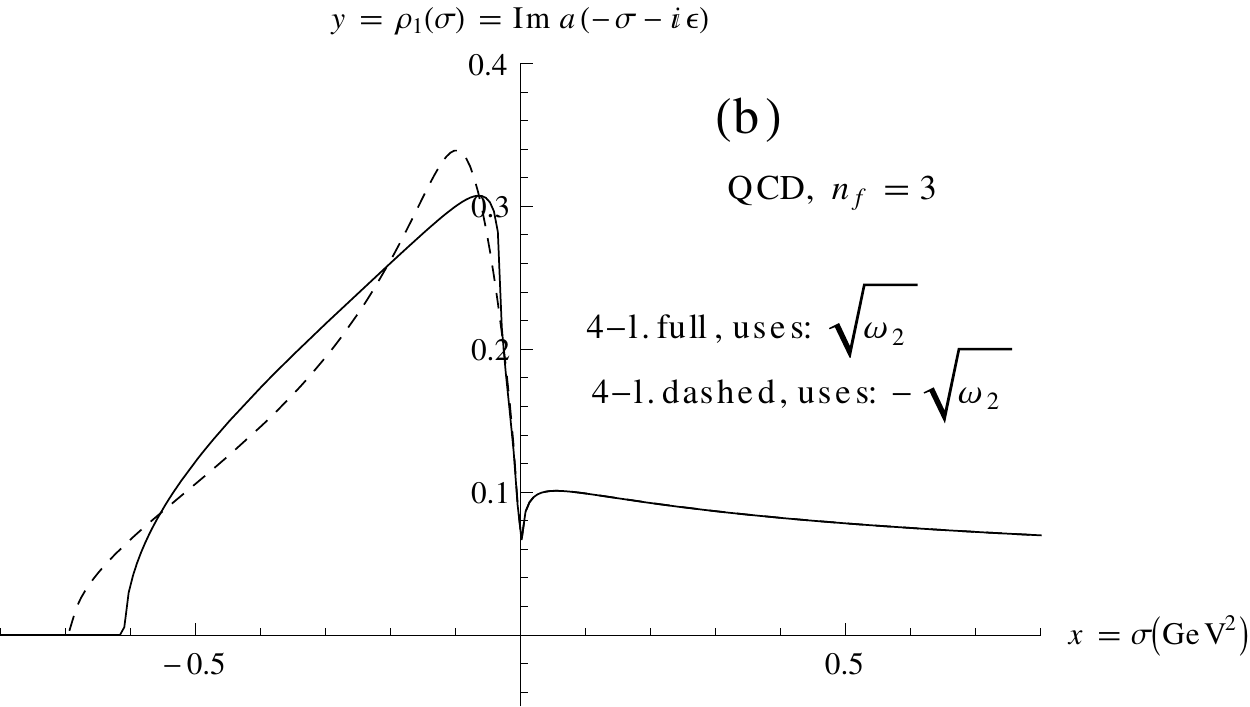}
\end{minipage}
\vspace{-0.4cm}
 \caption{\footnotesize  (a) The four-loop effective
running couplings $a^{(+)}$ (full line) and $a^{(-)}$ (dashed line),
at positive $Q^2$; (b) The discontinuity function 
$\rho_1(\sigma)= {\rm Im} a(Q^2=-\sigma - i \epsilon)$ 
as a function of $\sigma$ in these two cases.}
\label{aminus}
 \end{figure}

If we use in our effective four-loop case minus sign in front of $\sqrt{\omega_2}$
of eqs.~(\ref{choice}), i.e., the solution $a^{(-)}(Q^2)$, eq.~(\ref{am}), 
it turns out that the numerical results are very similar to the case
$a^{(+)}(Q^2)$, as it should be, except for at small values of $|Q^2|$. 
In fig.~\ref{aminus}(a) we compare,
at positive $Q^2$, the running coupling $a^{(-)}$ with the previously
presented coupling $a^{(+)}$. In fig.~\ref{aminus}(b), the discontinuity
functions $\rho_1(\sigma)$ are compared for the two cases.
The $\Lambda$ scale for $a^{(-)}$ is very high: $\Lambda\approx 0.560$ GeV,
compared to $\Lambda \approx 0.122$ GeV for $a^{(+)}$. Nonetheless, the 
Landau branching point is only a little higher: $Q_b \approx 0.832$ GeV
($Q_b \approx 0.781$ GeV for $a^{(+)}$). 

The branching point in the
case of $a^{(-)}$ is the branching point of the Lambert function
$W_{\mp 1}(z_b) = -1$ ($z_b=-1/e$; $Q_b^2 \approx 0.692 \ {\rm GeV}^2$), as argued
earlier in this Subsection, and the expression under the square root in
$a^{(-)}(Q^2)$ of eq.~(\ref{am}) is at this point positive.
On the other hand, the expression under the square root in $a^{(+)}$, 
at the point where $W=-1$, is negative; therefore, 
as argued earlier in the previous Subsection,
in this case the branching point of $a^{(+)}$ is determined by the point 
where the expression under the square root is zero 
($Q_b^2 \approx 0.610  \ {\rm GeV}^2$). 

In this context, we mention that 
the square root in eq.~(\ref{ap}) for $a^{(+)}$ should be evaluated
with caution, because at small $|Q^2|$ values the expression 
(``det'') under the
square root is complex and crosses the negative axis 
in the complex plane.
Specifically, at a fixed nonnegative argument of $\phi$ 
($0 \geq \phi \leq \pi$),
when $|Q^2|$ decreases from from the asymptotic region
($|Q^2| = + \infty$) along the ray towards the origin,
this expression 'det' varies continuously in the complex plane
from the 1st quadrant counterclockwise
in the following order: 1st $\to$ 2nd $\to$ 3rd $\to$ 4th 
(when $\phi <0$, 'det' travels in the clockwise direction,
since ${\rm det}(-\phi) = {\rm det}(\phi)^{\ast}$).
This continuity in the variation of the argument
of 'det' (between zero and $2 \pi$) 
must be reflected also in the square root ($\sqrt{{\rm det}}$). 
However, the numerical softwares usually assign
the values ${\rm Arg}({\rm det})$ ($\equiv \psi$) in the interval
$(- \pi, \pi)$, and such assignment would lead to
spurious discontinuities of $\sqrt{|{\rm det}|} \exp(i \psi/2)$
when 'det' crosses the negative semiaxis.
This means that the square root in eq.~(\ref{ap}) 
for $a^{(+)}$ must be implemented, for $Q^2 = |Q^2| \exp(i \phi)$ and
${\rm det} = |{\rm det}| \exp(i \psi)$, in any numerical evaluation, 
in the following way:
\begin{itemize}
\item
If $\phi \geq 0$: when $\psi \geq 0$ then $\sqrt{{\rm det}} = \sqrt{|{\rm det}|} \exp(i \psi/2)$; when $\psi < 0$ then $\sqrt{{\rm det}} = \sqrt{|{\rm det}|} \exp(i (\psi + 2 \pi)/2)$.
\item
If $\phi < 0$: when $\psi \leq 0$ then $\sqrt{{\rm det}} = \sqrt{|{\rm det}|} \exp(i \psi/2)$; when $\psi > 0$ then $\sqrt{{\rm det}} = \sqrt{|{\rm det}|} \exp(i (\psi - 2 \pi)/2)$.
\end{itemize}

In the case of $a^{(-)}$ this rule for the evaluation of 
$\sqrt{{\rm det}}$ does not lead to any change of the result when 
compared to the naive evaluation of $\sqrt{{\rm det}}$,
as already argued in the previous subsection.

The presented effective four-loop numerical evaluations 
based on our formulas (\ref{ap})-(\ref{am})
were cross-checked by performing the numerical RGE integration in
the complex plane of $\ln(Q^2/\mu^2_{\rm in})$, applying the same
initial condition $a=0.07245$ at $Q^2=\mu^2_{\rm in}$. The formulas
(\ref{ap})-(\ref{am}) are, of course, numerically much more efficient
than the numerical RGE integration in the complex plane. This
practical usefulness is, in fact, the main motivation for 
deriving and presenting the explicit solutions 
eqs.~(\ref{ap})-(\ref{am}) in the context of QCD.

\section{Effective five-loop case}
\label{sec:5l}

\subsection{Five-loop ansatz for $f(u)$}
\label{subsec:5lansf}

We take $f(u)$ in the form 
\begin{eqnarray}
 f(u) = u + a_0 + \frac{a_1}{u} + \frac{a_2}{u^2}, \label{AA1}
\end{eqnarray}
where $a_0,$  $a_1$ and $a_2$ are arbitrary (real) numbers. We will show 
how to reproduce the chosen coefficients of the $\beta$-function   
up $\sim a^6$
\begin{eqnarray*}
\beta(a) = \beta_0 a^2~(1+c_1 a  + c_2 a^2 + c_3 a^3 + c_4 a^4 ) +
{\cal O}(a^7) \ .
\end{eqnarray*}
Thus, we consider the $\beta$-function ansatz as in the previous section 
\begin{eqnarray*}
\beta(a) = - \frac{A a^2}{2}\frac{B/f_u'(u)}{1 - 1/f(u)}, 
\end{eqnarray*}
where $f(u)$ is now given in eq.~(\ref{AA1}). 
In this effective ``five-loop'' case, we repeat the same procedure
as was applied in the effective ``four-loop'' case in
subsec.~\ref{subsec:4lansf}, except that now the expansion of
$1/f^{'}(u)$ must be performed up to $\sim a^4$. Equation (\ref{be0}) and the 
first two relations in eq.~(\ref{ajcj1}) are reproduced. However,
the third relation in eq.~(\ref{ajcj1}) is modified due to the
presence of $a_2$,
\begin{equation*}
( (a_0-1)^2 + 2 a_2) (B/2)^3 = c_3 \ ,
\label{a2c3}
\end{equation*}
and, additionally, we get a relation involving $c_4$
\begin{equation*}
a_2   + a_1(a_0-1)   - (a_0 -1)^3  + a_1^2  = c_4/c_1^4  \equiv \omega_3.
\label{a2c4}
\end{equation*}
Thus, we obtain
\begin{equation}
B = 2 c_1, \quad A=\beta_0/c_1 \ ,
\label{AB}
\end{equation}
and the other three relations take the form
\begin{eqnarray*}
a_1 - a_0 + 1 = c_2/c_1^2 \equiv \omega_1,\\
(a_0 -1)^2 + 2a_2 = c_3 /c_1^3  \equiv \omega_2,\\
a_2   + a_1(a_0-1)   - (a_0 -1)^3  + a_1^2  = c_4/c_1^4  \equiv \omega_3.
\end{eqnarray*}
As a consequence, the system of the first and the second equation is 
\begin{eqnarray}
a_1 =  \omega_1 + a_0 - 1,  \nonumber\\
a_2 = (\omega_2 - (a_0 -1)^2)/2.  
\label{A1A2}
\end{eqnarray}
After substituting this in the last equation we obtain  
\begin{eqnarray*}
( \omega_2 - (a_0 -1)^2)/2  +  (\omega_1 + a_0 - 1)(a_0-1) \\
- (a_0 -1)^3  + (\omega_1 + a_0 - 1)^2  = \omega_3,
\end{eqnarray*}
which can be transformed to a form of cubic equation for $a_0$ 
\begin{eqnarray}
- (a_0 -1)^3 + \frac{3}{2}(a_0 -1)^2  +  3\omega_1 (a_0-1) 
= \omega_3 -  \omega^2_1 - \omega_2/2.  
\label{naf}
\end{eqnarray}
This is the cubic equation that has either one real or three real solutions. 
The knowledge of $a_0$ gives the values of $a_1$ and $a_2$ 
via eqs.~(\ref{A1A2}).

\subsection{Cardano solution to cubic equation}
\label{subsec:card}

The solution to cubic equation has been found by Cardano 
\cite{Cardano1, Cardano2}. Any cubic equation can be 
transformed to the form without the second power of the unknown variable
$x$
\begin{eqnarray}
x^3 + px + q = 0.
\label{cub}
\end{eqnarray}
The result for $x$  is given by the Cardano formula 
\begin{eqnarray}
x = \sqrt[3]{-\frac{q}{2} + \sqrt{\frac{q^2}{4} + \frac{p^3}{27}}} 
+ \sqrt[3]{-\frac{q}{2} - \sqrt{\frac{q^2}{4} + \frac{p^3}{27}}}.  
\label{sol} 
\end{eqnarray}
The type of solution is determined by the sign of the value of $\tau$
\begin{eqnarray}
\tau \equiv \frac{q^2}{4} + \frac{p^3}{27}.  \label{5.4.tau}
\end{eqnarray}
If $\tau$ is real and positive, 
there is one real root and two complex ones;
if $\tau < 0$, there are three distinct real roots. 
The case $\tau = 0$ corresponds to the three real roots but two 
of them coincide. 
The expression (\ref{sol}) represents several (nine) possible roots.
The true roots of eq.~(\ref{cub}) are only those for which the
product of the two terms in eq.~(\ref{sol}) is equal to $-p/3$
\begin{equation}
\sqrt[3]{-\frac{q}{2} + \sqrt{\frac{q^2}{4} + \frac{p^3}{27}}} 
 \sqrt[3]{-\frac{q}{2} - \sqrt{\frac{q^2}{4} + \frac{p^3}{27}}}
= - \frac{p}{3} \ .
\label{restr}
\end{equation}
This appears to hold even in the most general case when $p$ and $q$ 
(and thus $\tau$) are nonreal complex numbers.

\subsection{Equation for $a_0$}
\label{subsec:solf5l}

We write eq.~(\ref{naf}) in the form 
\begin{eqnarray}
x^3 - \frac{3}{2}x^2 - 3\omega_1 x + 
(\omega_3 -  \omega^2_1 - \omega_2/2)  = 0, 
\label{eqa0o1}
\end{eqnarray}
where $x= a_0 -1,$ and transform it to the form 
without the quadratic terms
\begin{eqnarray}
\left(x-\frac{1}{2}\right)^3 - \left(\frac{3}{4} + 
3\omega_1\right)\left(x-\frac{1}{2}\right) 
- \frac{1}{4} + (\omega_3 -  \omega^2_1 - \omega_2/2) 
- \frac{3}{2}\omega_1 = 0.
\label{eqa0o2}
\end{eqnarray}
Thus, to obtain $a_0$ we can use the solution (\ref{sol}) for the variable 
$\left(x-\frac{1}{2}\right) \equiv (a_0 - 3/2)$, by substituting 
\begin{equation}
p = - \frac{3}{4} - 3\omega_1, \quad
q = - \frac{1}{4} + \omega_3 -  \omega^2_1 - \omega_2/2 - \frac{3}{2}\omega_1
\ .
\label{eqa0o2b}
\end{equation}
Since in this case all the coefficients are real, 
we should obtain at least one real root.  
After finding $a_0$, we obtain $a_1$ and $a_2$ from eqs.~(\ref{A1A2}).

\subsection{The inverse function for $f(u)$}
\label{subsec:invf5l}

Knowing explicitly the coefficients 
$a_0,$ $a_1$ and $a_2$ in the ansatz (\ref{AA1}), we have
\begin{eqnarray}
-W(z) = u + a_0 + \frac{a_1}{u} + \frac{a_2}{u^2},  
\label{type2}
\end{eqnarray}
where $u = 2/(B a) = 1/(c_1 a)$, and $z$ is given in eq.~(\ref{zz}).
We rewrite eq.~(\ref{type2}) in the form 
\begin{eqnarray*}
u^3 + (a_0 + W) u^2 + a_1 u + a_2 = 0,
\end{eqnarray*}
that is again the usual cubic equation. 
We can rewrite it in the form
\begin{eqnarray}
\lefteqn{
\left( u + (a_0 + W)/3 \right)^3 + 
\left( a_1 - (a_0 + W)^2/3 \right) \left( u + (a_0 + W)/3 \right) }
\nonumber\\
&& -  a_1 (a_0 + W)/3 + 2(a_0 + W)^3/27 + a_2 = 0 \ , 
\label{eqaQ25l}
\end{eqnarray}
i.e., we eliminate the term with the second power. Thus, 
to solve for the running coupling $a(Q^2) \equiv 1/(c_1 u)$,
we can use the Cardano solution (\ref{sol}) where
\begin{eqnarray}
x = x(z) &= & u + \frac{1}{3}\left( a_0 + W(z) \right) =
\frac{1}{c_1 a(Q^2)} + \frac{1}{3}\left( a_0 + W(z) \right) 
\ \Rightarrow
\nonumber\\
a(Q^2) & = & \frac{1}{c_1} \frac{1}{ \left[
x(z) - \left( a_0 + W(z) \right)/3 \right] } \ ,
\label{xa}
\\
p(z) &=& - \frac{1}{3} \left( a_0 + W(z) \right)^2 + \omega_1 + a_0 - 1 \ , 
\label{pa0}
\\
q(z) &= & \frac{2}{27} \left( a_0 + W(z) \right)^3 -  
\frac{\omega_1 + a_0 - 1}{3} \left( a_0 + W(z) \right) 
+ \frac{\omega_2 - (a_0 -1)^2}{2} \ .
\label{qa0}
\end{eqnarray}
As in sec.~\ref{sec:4l}, $W(z) = W_{\mp 1}(z)$ is again the Lambert function,
with $z$ defined in eq.~(\ref{zz}), and $W=W_{-1}(z)$ when
$0 \leq \phi < \pi$, and $W=W_{+1}(z)$ when $-\pi \leq \phi < 0$ (where:
$Q^2 = |Q^2| e^{i \phi}$). Now, for a general complex $z$ (or: $Q^2$),
the coefficients $p(z)$ and $q(z)$ are complex numbers, too, as are the roots
$x(z)$ and thus $a(Q^2)$. The restriction (\ref{restr}) is valid also
in this general complex case.

\subsection{Cut structure and analyticity}
\label{subsec:cut5l}

An advantage of the effective five-loop solution, 
eqs.~(\ref{xa})-(\ref{qa0}) and (\ref{sol}), 
in comparison with the four-loop solution, eqs.~(\ref{ap}) and (\ref{am}),
is that the cubic equation (\ref{naf}) for the coefficient $a_0$
always has at least one real root. 
This means that we do not obtain any restriction on the
sign of the RSch coefficients $c_j$ ($j=2,3,4$) imposed by the
reality of the running coupling $a(Q^2)$ at large positive $Q^2$,
in contrast to the effective four-loop case (where $c_3<0$ leads
to a nonreal $a_0$ and nonreal $a(Q^2)$ at large positive $Q^2$).

As has been shown in the previous section, dedicated to the 
effective four-loop case, the cut structure of $a(Q^2)$
in the plane of complex $z$ (or: of complex $Q^2$) has two origins.
The first origin of the cuts is the multivalued nature 
of the Lambert W function. In the effective five-loop case,
the corresponding cuts in $z$-plane repeat completely their analogs 
in the effective four-loop case; the corresponding Riemann surface 
remains unchanged, and also the phase relation 
caused by the power-like relation between the complex variable $z$ 
and the complex variable $Q^2$, eq.~(\ref{zz}), remains the same.

The second origin for the cuts in the complex $z$-plane is that 
the running coupling $a(Q^2)$
is a composite function of the Lambert function, that composite
function having multivalued nature, too. 
In the effective five-loop case of this section, the cut structure 
in the $z$-plane is different from the corresponding cut structure 
in the effective four-loop case (\ref{5.4.FLCW})
since the map from the $W$-partitions to the running coupling has a more 
complicated structure, eqs.~(\ref{xa})-(\ref{qa0}) and (\ref{sol}).
To identify the start of the cut in the complex $W$-plane, we have  
to analyze the roots of the quantity $\tau$, eq.~(\ref{5.4.tau}),
i.e., roots of the sixth power polynomial 
described in terms of the variable $a_0+W(z)$,
cf.~eqs.~(\ref{pa0})-(\ref{qa0}).
At present, it is impossible  
to find analytic expressions for the roots of polynomials of 
power higher than four.
However there are many numerical tools
to perform that analysis (see the next subsection). 
In addition, the possible cuts due to the cubic roots in (\ref{sol}) should 
be taken into account. These cubic roots can produce 
new cuts in the partitions of the complex plane of the 
composite variable $\sqrt{\tau}$, which would 
be transformed into new cuts in the $W$-plane, 
and finally into new cuts in the $z$-plane (and $Q^2$-plane). 
Such a recursive mapping of the cuts 
is the simplest way to analyze the cut structure
of the running coupling in the complex $z$-plane 
(and $Q^2$-plane), especially for the evaluation of the
starting (branching) points of the cuts.   
 
From the point of view of the analytic evaluation, a simpler representation
for the solutions exists in terms of trigonometric functions \cite{Smirnov}. 
It is not a universal representation 
as the representation of eq.~(\ref{sol}).
For example, for real negative $\tau$ and $p$, 
the trigonometric representation of eq.~(\ref{sol}) is
\begin{eqnarray*}
x = 2\sqrt[3]{r}\cos \left( {\frac{\phi + 2k\pi}{3}} \right), 
~~~~ k = 0,1,2, \\
r = \sqrt{-\frac{p^3}{27}}, ~~~~ \cos\phi = -\frac{q}{2r}.
\end{eqnarray*}
It is a more helpful form of solution for the analysis of the roots;
but it is less helpful for the analysis of cuts in 
the complex $z$-plane, since the form of the representations 
is not power-like. Thus, we prefer 
the original form (\ref{sol}) of the Cardano solution 
for the numerical evaluation of the cuts in the complex $z$-plane.

\subsection{Numerical evaluation of analytic formulas}
\label{subsec:num5l}

Similarly as we did in subsec.~\ref{subsec:num4l} in the effective
four-loop case, we evaluate now the analytic formulas for the
running coupling $a(Q^2)$ in the effective five-loop case, in
QCD with $n_f=3$ and in ${\overline {\rm MS}}$ scheme. 
In contrast to the coefficients $c_2$ and $c_3$, the exact five-loop
coefficient $c_4$ in ${\overline {\rm MS}}$ has not been obtained yet in
the literature. However, Pad\'e-related estimates, ref.~\cite{Ellis:1997sb}
give the value $c_4 \approx 123.701$ at $n_f=3$. 
We use this value in our formulas.
Equation (\ref{eqa0o2}) for the coefficient $a_0$
has only one real solution, $a_0=-1.19666$. The
running coupling $a(Q^2)$ is obtained by using the formula (\ref{sol}) in
conjunction with the formulas (\ref{eqaQ25l}) and (\ref{xa})-(\ref{qa0}),
and the relation (\ref{zz}). The same initial condition as in 
subsec.~\ref{subsec:num4l} is applied, i.e., eq.~(\ref{ain}).
In the effective five-loop case, this condition gives
$\Lambda \approx 0.621 {\rm GeV}^2$. 

At large values of $|Q^2|/\Lambda^2 \equiv u$ ($ = |z|$),
i.e., in the asymptotic freedom regime, 
this gives us the correct real solution for $a(Q^2)$ unambiguously. However,
when we move in the $Q^2$-complex plane along a fixed ray $\phi = {\rm const}$
(where $Q^2 = |Q^2| e^{i \phi} = u \Lambda^2 u e^{i \phi}$) toward the origin, 
the expression under the square root in eq.~(\ref{sol}) 
\begin{equation}
{\rm Det}_2(u,\phi)  \left( = \tau(p,q) \right) \equiv
\frac{1}{4} \left(q(u,\phi)\right)^2 + \frac{1}{27} \left( p(u,\phi) \right)^3 
\label{Det2}
\end{equation}
changes the argument $\psi_2$ of ${\rm Det}_2 = |{\rm Det}_2| \exp(i \psi_2)$
continuously, and this behavior depends crucially
on whether the (nonnegative fixed) $\phi$ is below or above a threshold
angle $\phi_{\rm thr}$. Here, the threshold angle is determined by
the condition
\begin{equation}
{\rm Det}_2(u_{\rm thr},\phi_{\rm thr}) = 0  \ \Rightarrow
\phi_{\rm thr} \approx 0.0507, \ u_{\rm thr} \approx 0.314 \ .
\label{Det2thr}
\end{equation}
\begin{figure}[htb] 
\centering\includegraphics[width=100mm]{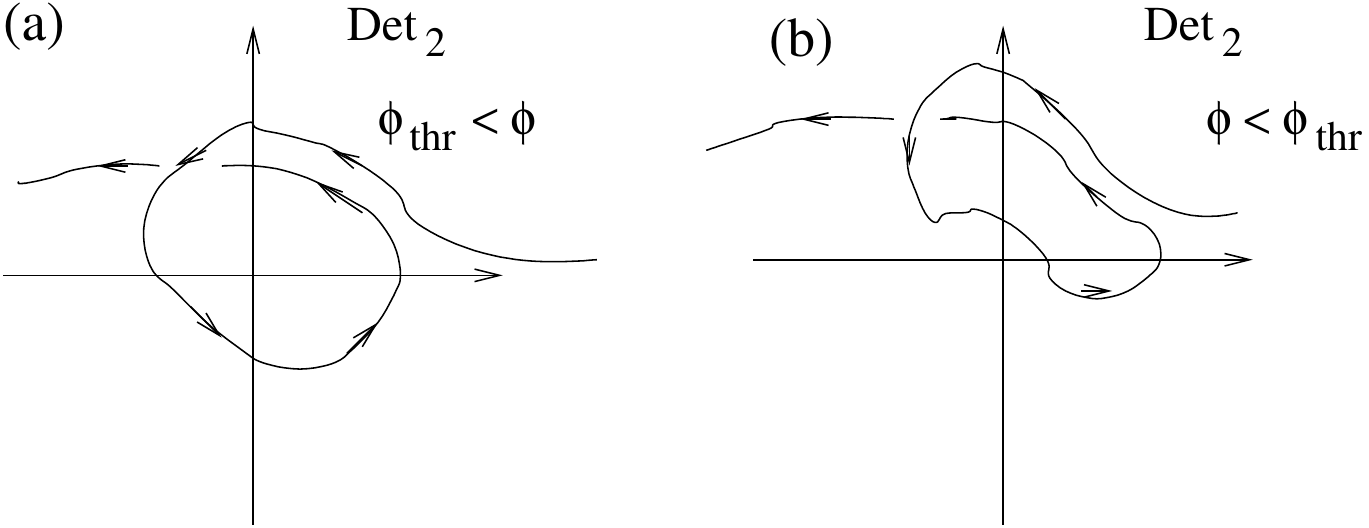}
\vspace{-0.4cm}
 \caption{\footnotesize  The variation of ${\rm Det}_2(u,\phi)$ in the
complex plane, at fixed nonnegative $\phi$ ($\equiv arg(Q^2)$), 
when $u$ ($\equiv |Q^2|/\Lambda^2$) is decreasing toward zero: 
(a) when $\phi_{\rm thr} < \phi$, the path encircles the origin;
(b) when  $\phi < \phi_{\rm thr}$, the path avoids encircling the origin.}
\label{FigDet2}
 \end{figure}
It turns out that,
at fixed nonnegative $\phi$ and when $u \equiv |Q^2|/\Lambda^2$ decreases 
toward zero,
the argument $\psi_2$ of ${\rm Det}_2$ varies 
within the interval $(0, 3 \pi)$ if $\phi_{\rm thr} < \phi$,
and in the interval $(-\pi/2,+\pi)$ if ($0 \leq$) $\phi < \phi_{\rm thr}$ --
see figs.~\ref{FigDet2}(a) and (b). The square root of ${\rm Det}_2$
must be calculated in the way that reflects the continuous change of
$\psi_2$ during the movement along the ray, i.e.,
$\sqrt{{\rm Det}_2} = \sqrt{|{\rm Det}_2|} \exp(i \psi_2/2)$.\footnote{
We recall that in the softwares, e.g. in Mathematica \cite{Math8},
the arguments of the numerically evaluated complex numbers
vary in the restricted interval $]-\pi,+\pi]$, i.e.,
unfortunately, no distinction is made between the 
1st and the 5th quadrant, etc.}

The same care must be taken when evaluating the third roots in
the solution (\ref{sol}), i.e., the third roots of
\begin{equation}
{\rm Det}_{3 \pm}(u,\phi)  \equiv  -\frac{1}{2} q(u,\phi)
\pm \sqrt{{\rm Det}_2(u,\phi)} \ .
\label{Det3pm}
\end{equation}
Again, the behavior along the $Q^2$-rays changes drastically when
$\phi$ passes the threshold angle $\phi_{\rm thr}$. 
\begin{figure}[htb] 
\centering\includegraphics[width=100mm]{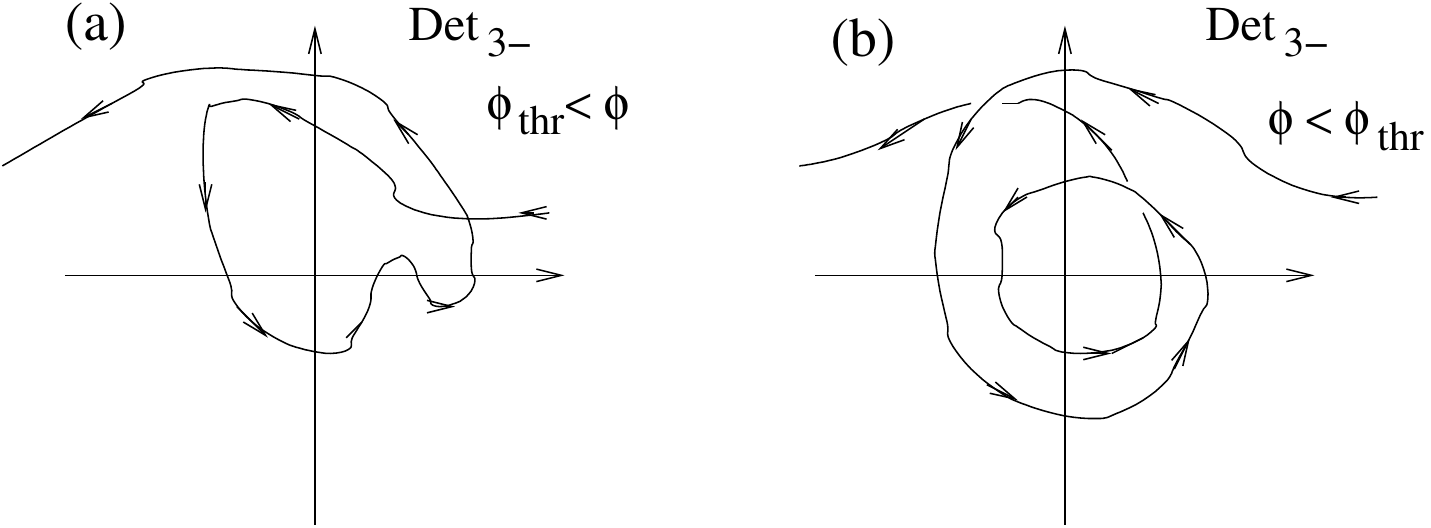}
\vspace{-0.4cm}
 \caption{\footnotesize  The same as in fig.~\ref{FigDet2},
but now for the expression ${\rm Det}_{3-}(u,\phi)$ given 
in eq.~(\ref{Det3pm}).}
\label{FigDet3m}
 \end{figure}
\begin{figure}[htb] 
\centering\includegraphics[width=100mm]{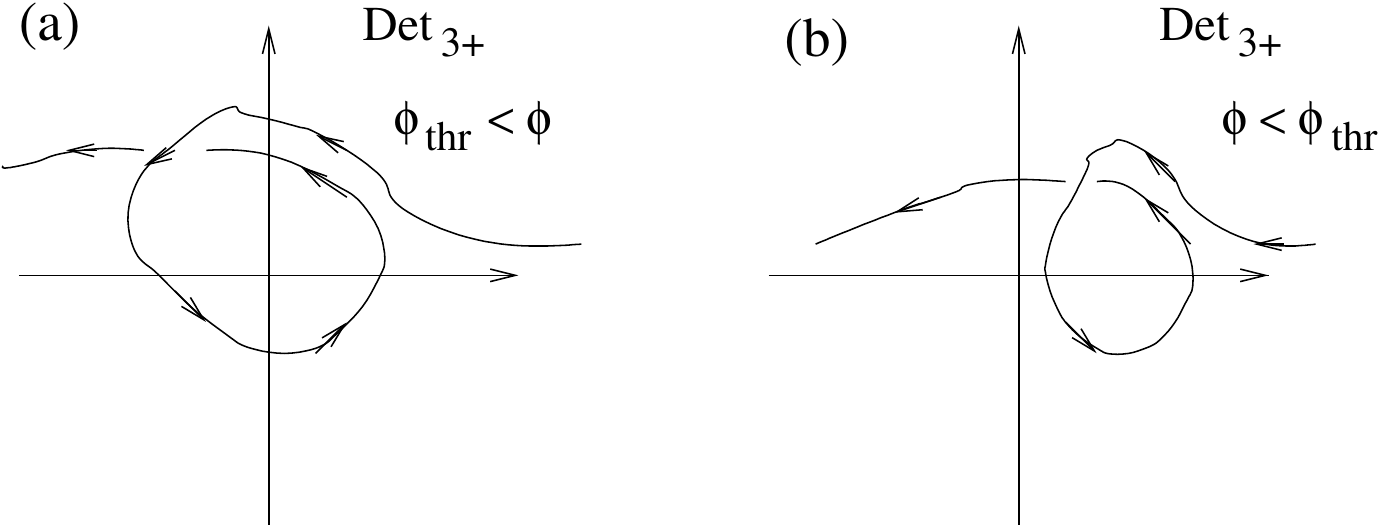}
\vspace{-0.4cm}
 \caption{\footnotesize The same as in fig.~\ref{FigDet2},
but now for the expression ${\rm Det}_{3+}(u,\phi)$ given 
in eq.~(\ref{Det3pm}).} 
\label{FigDet3p}
 \end{figure}

The argument $\psi_{3-}$
of ${\rm Det}_{3-}$ varies in the interval $(0, +3 \pi)$ if $\phi_{\rm thr} < \phi$,
and in $(0, +5 \pi)$ if ($0 \leq$) $\phi < \phi_{\rm thr}$ -- 
cf.~figs.~\ref{FigDet3m}(a) and (b). 
The argument $\psi_{3+}$ of ${\rm Det}_{3+}$ varies 
in the interval $(0, +3 \pi)$ if $\phi_{\rm thr} < \phi$,
and in $(-\pi/2, + \pi)$ if ($0 \leq$) $\phi < \phi_{\rm thr}$ -- 
cf.~figs.~\ref{FigDet3p}(a) and (b). The third roots in the solution
(\ref{sol}), whose sum gives us the running complex $a(Q^2)$, 
must reflect the continuous change of
$\psi_{3 \pm}$ during the movement along the rays, i.e.,
in the evaluation we must implement:
$({\rm Det}_{3 \pm})^{1/3} = |{\rm Det}_{3 \pm}|^{1/3} \exp(i \psi_{3 \pm}/3)$.

For the negative $\phi$ ($- \pi \leq \phi < 0$), the evaluation of
$a(Q^2)$ is then implemented easily, using the relation $a(Q^{2 \ast})
= a(Q^2)^{\ast}$.

In this way, the correct roots are chosen from the plethora of possible
roots given by the expression (\ref{sol}), for all complex $Q^2$ down to
$Q^2 \to 0$. In the Appendix we specify, for the case of QCD in
the  ${\overline {\rm MS}}$ scheme with $n_f=3$ as an example, how to
implement in practice (in software) the calculation of the angles 
$\psi_2$ and $\psi_{3 \pm}$, and thus the evaluation of $({\rm Det_2})^{1/2}$
and $({\rm Det}_{3 \pm})^{1/3}$.
We checked that the solutions obtained in this way satisfy
eq.~(\ref{cub}) with $p$ and $q$ given in eqs.~(\ref{pa0})-(\ref{qa0}),
and the constraint (\ref{restr}). 

\begin{figure}[htb] 
\centering\includegraphics[width=100mm]{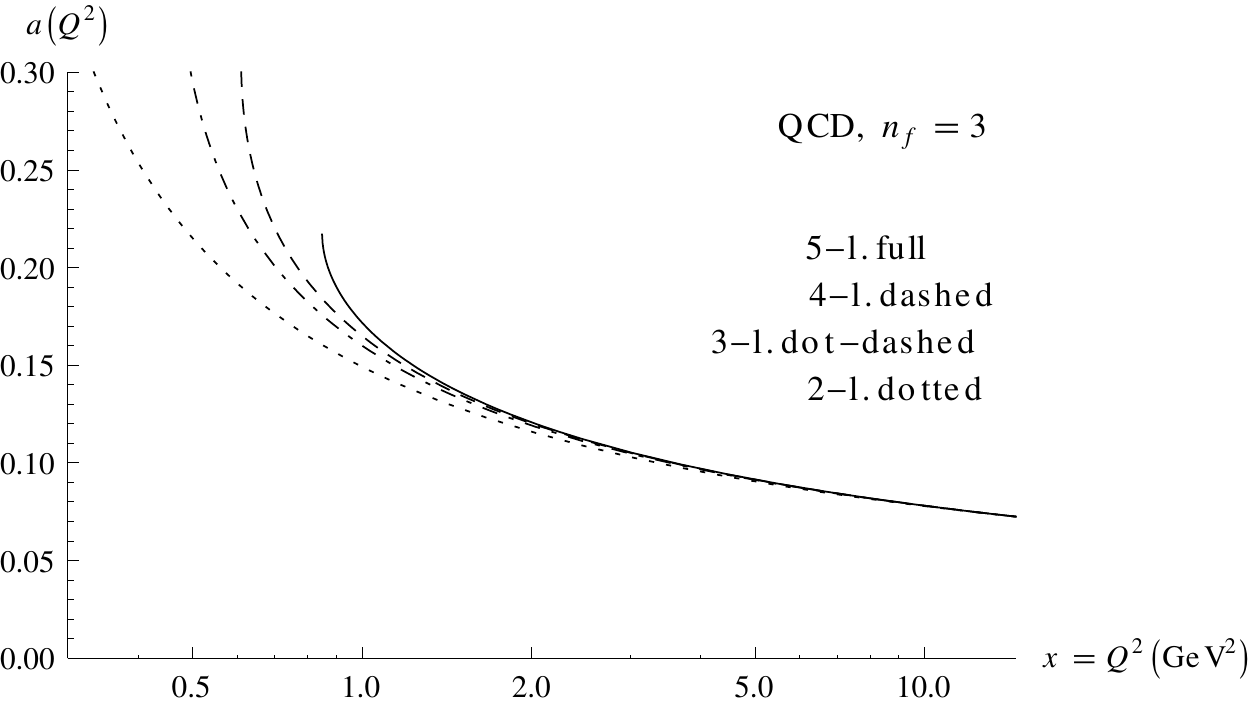}
\vspace{-0.4cm}
 \caption{\footnotesize  The effective five-loop running coupling
$a^{(+)}(Q^2)$ at positive $Q^2$. For comparison, the effective four- and
three-loop coupling, and the two-loop coupling, are included.}
\label{aQ25l}
 \end{figure}
In fig.~\ref{aQ25l} we present the effective five-loop 
running coupling, evaluated in the aforementioned way,
at positive $Q^2$. The lower-loop couplings are included, for comparison.
The figure is analogous to fig.~\ref{aQ2}, but now includes the
effective five-loop case. We recall that all the couplings are adjusted
to the same initial value at $Q^2 = \mu_{\rm in}^2 \approx 14.516 \ {\rm GeV}^2$,
eq.~(\ref{ain}).

\begin{figure}[htb] 
\begin{minipage}[b]{.49\linewidth}
\centering\includegraphics[width=70mm]{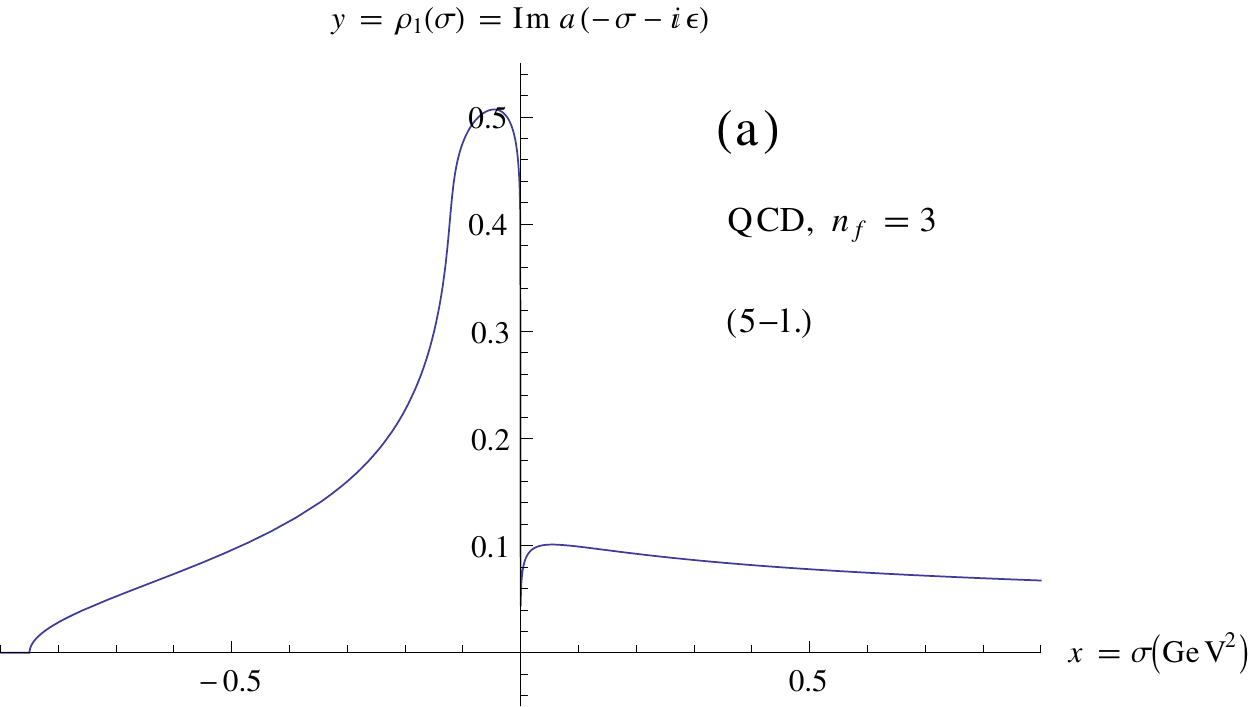}
\end{minipage}
\begin{minipage}[b]{.49\linewidth}
\centering\includegraphics[width=70mm]{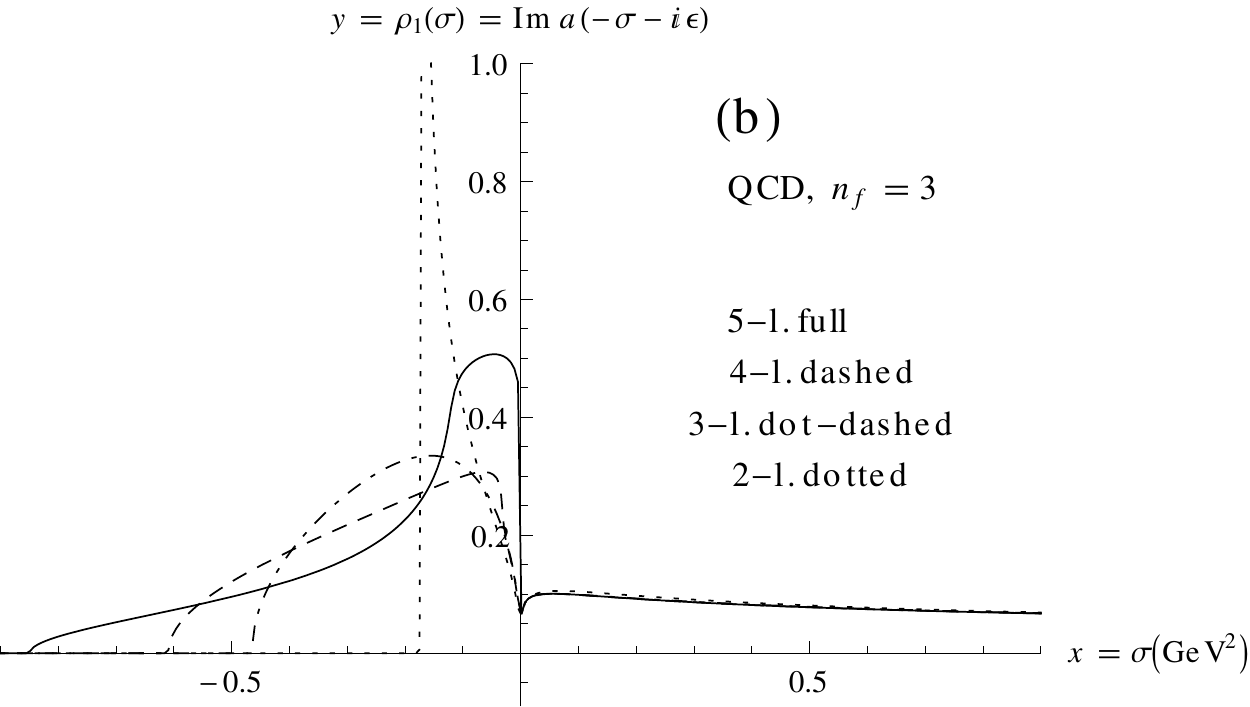}
\end{minipage}
\vspace{-0.4cm}
 \caption{\footnotesize  (a) The effective five-loop discontinuity function 
$\rho_1(\sigma)= {\rm Im} a^{(+)}(Q^2=-\sigma - i \epsilon)$ 
as a function of $\sigma$; (b) for comparison, $\rho_1(\sigma)$
for the effective four- and three-loop case, and for the two-loop case, 
are included.}
\label{rho15l}
 \end{figure}
In fig.~\ref{rho15l}(a) we present the effective five-loop
discontinuity function $\rho_1(\sigma) = {\rm Im} a(-\sigma - i \epsilon)$.
The branching point where the (Landau) cut starts is at
$Q_b^2  \approx 0.849 \ {\rm GeV}^2$ ($\sigma_b = - Q_b^2$), 
i.e., at $z = -1/e$ where the Lambert
function $W_{-1}(z)$ has the value $-1$ and has a branching point.
In fig.~\ref{rho15l}(b), this discontinuity function is compared
with those of the effective four- and three-loop case
and the two-loop case. The aforementioned branching point can be inferred
also from fig.~\ref{aQ25l}, where we see that the (effective five-loop)
coupling achieves a finite value $a_b \approx 0.218$ 
at the branching point.\footnote{If we use for
$\beta(a)$ the power series in $a$ truncated at the
five loop-level, and the same $a_{\rm in}$ value 
as in eq.~(\ref{ain}), the numerical integration of the RGE 
gives for the branching point the value $Q_b^2 \approx 0.496 \ {\rm GeV}^2$; 
the value $a(Q_b^2)$ is infinite in such a case.}

\begin{figure}[htb] 
\begin{minipage}[b]{.49\linewidth}
\centering\includegraphics[width=75mm]{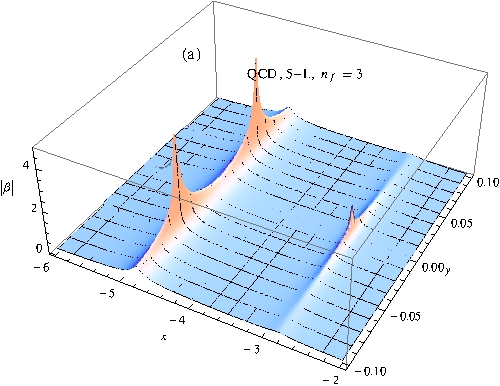}
\end{minipage}
\begin{minipage}[b]{.49\linewidth}
\centering\includegraphics[width=75mm]{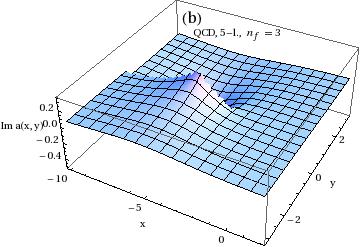}
\end{minipage}
\vspace{-0.4cm}
 \caption{\footnotesize  (a) The absolute value of the beta function
$\beta(a(Q^2))$, for complex $Q^2$ ($= \mu_{\rm in}^2 \exp(x) \exp(i y)$) near the
origin; the three starting (branching) points of Landau cuts are
visible as three peaks; (b) imaginary part of $a(Q^2)$ in the entire
$Q^2$ complex plane.}
\label{Figbetv5l}
 \end{figure}

In the presented ${\overline {\rm MS}}$ effective five-loop case (with $n_f=3$), 
there is another branching point, namely the one given in
eq.~(\ref{Det2thr}), as already mentioned in the
previous subsection. It corresponds to the value
$|Q^2_{\rm thr}| \approx 0.122 \ {\rm GeV}^2$ and 
$\phi = \pm \phi_{\rm thr} \approx \pm 0.0507$.
This branching point is quite close to the origin, but is off the
real axis. In order to see this starting point of nonanalyticity
more easily, we present in fig.~\ref{Figbetv5l}(a) the 
function $|\beta(a(Q^2))|$ at complex $Q^2$ not far from the origin,
with axes $x=\ln(Q^2/\mu_{\rm in}^2)$ (note: $\mu_{\rm in}^2 \approx 14.516 \ {\rm GeV}^2$)
and $y = \phi$. In the figure we clearly see two peaks at $x \approx -4.79$
($|Q^2| \approx 0.122 \ {\rm GeV}^2$) and $\phi \approx \pm \phi_{\rm thr}$. Yet another
peak is visible at $x \approx -2.84$ and $\phi=0$, representing the
aforementioned $Q_b^2 \approx 0.849 \ {\rm GeV}^2$ branching point that is
simultaneously a branching point of $W_{-1}(z)$.

In fig.~\ref{Figbetv5l}(b) we present the imaginary part of $a(Q^2)$ in
the (almost) entire complex plane. 
At $y=0$ the usual discontinuity function $\rho_1$
is contained in this figure.

\section{Conclusion}
\label{sec:concl}

In this paper we started with the derivation of an analytic formula
for the solution of the renormalization group equation (RGE) for the
coupling parameter in the supersymmetric model of 
Novikov et al.~(NSVZ, \cite{Novikov:1983uc}). 
We extended this analysis, by deriving analytic formulas for solutions to 
a class of RGE's for the QCD coupling parameter $a(Q^2) \equiv \alpha_s(Q^2)/\pi$.
The class of beta functions $\beta(a)$ in these RGE's is of the form
\begin{equation}
\frac{d a(Q^2)}{d \ln Q^2} = - \beta_0 a^2 
\frac{ \left( 1/f_u^{'}(u) \right) }
{ \left( 1 - 1/f(u) \right) }{\bigg |}_{u=1/(c_1 a)} \ ,
\label{purRGEs}
\end{equation}
where $\beta_0$ and $c_1 = \beta_1/\beta_0$ are the (universal) first two
coefficients in the power expansion of $\beta$ function eq.~(\ref{beta1}), 
and $f(u)$ is a function of the following form:
\begin{equation}
f(u) = u + a_0 + \sum_{j=1}^n \frac{a_j}{u^j} \ .
\label{fans}
\end{equation}
We found that the solution $a(Q^2)$, for general complex $Q^2$,
can be written in the following simple (but implicit) form:
\begin{equation}
f(u)|_{u=1/(c_1 a(Q^2))} = 
- W_{\mp 1} \left( - (\Lambda^2/Q^2)^{\beta_0/c_1} \right) \ ,
\label{implsol}
\end{equation}
where $W_{\mp 1}$ are two partitions (branches) of the Lambert function 
($W_{-1}$ when ${\rm Im} Q^2 \geq 0$, $W_{+1}$ when ${\rm Im}(Q^2)<0$), 
and the scale $\Lambda$ is fixed by
an initial condition [e.g., the condition (\ref{ain})]. 

We showed that
the $(n+1)$ real parameters $a_j$ ($j=0,\ldots,n$) in function $f(u)$ can
be adjusted so that the power expansion of $\beta(a)$ reproduces
the $(n+3)$-loop polynomial beta function in any chosen renormalization
scheme (RSch), i.e., for any chosen values of the RSch parameters
$c_2, \ldots, c_{n+2}$ in the expansion (\ref{beta1}).
In order to obtain an analytic (i.e., explicit) formula for
$a(Q^2)$, the polynomial-type of relation (\ref{implsol}) 
has to be solved. This we did explicitly in the (effective)
four-loop case ($n=1$) and in the effective five-loop case ($n=2$).
The (effective) three-loop case ($n=0$), i.e., the case of the
beta function of eq.~(\ref{beta3lGardi}), was solved in
ref.~\cite{Gardi:1998qr} (their sec.~4).

We discussed in detail the (non)analyticity structure of the RGE
solution in the complex $Q^2$ plane. We presented numerical evaluation
of the obtained formulas in the case of the effective four-loop and
the effective five-loop $\beta$ (i.e., for $n=1, 2$) in the
${\overline {\rm MS}}$ RSch and with the number of active quark flavors
$n_f=3$.

It is, in principle, 
possible to go even further, to $n=3$ (effective six-loop case).
In such a case eq.~(\ref{implsol}) becomes a quartic equation
in $u$ ($\equiv 1/(c_1 a)$), i.e., the highest order polynomial equation
for which an analytic (explicit) formula exists: Ferrari formula.
In any case, the (effective) five-loop formula found in
sec.~\ref{sec:5l} is already 
a good approximation at sufficiently high $|Q^2|$ 
(e.g., at $|Q^2| > 2 \ {\rm GeV}^2$, cf.~fig.~\ref{aQ25l}).
Any numerical evaluation of perturbative results at five-loop order 
can be performed by using the (effective) five-loop formula for
$a(Q^2)$ found in this paper.

\begin{acknowledgments}
\noindent
The work of G.C.  was supported in part by Fondecyt (Chile) grant  \#1095196 and Anillos Project ACT119.
The work of I.K. was supported in part by Fondecyt (Chile) grants  \#1040368, \#1050512 and by DIUBB grant (UBB, Chile) \#102609. 
We are grateful to Tim Jones for his comments on the first version of the present paper, for attracting our attention to 
the second entry of Ref. \cite{Novikov:1983uc} 
and for his interest to this research result.

\end{acknowledgments}

\appendix

\section{Implementation of arguments in the effective five-loop case}
\label{App}

In the effective five-loop case of sec.~\ref{sec:5l}, for the
specific case of QCD with $n_f=3$ in ${\overline {\rm MS}}$ scheme,
the results of the numerical evaluation of the analytic formula 
for the running coupling $a(Q^2)$ (with $Q^2$ in general complex) 
were presented in subsec.~\ref{subsec:num5l}. For such evaluations,
we need to know in practice how to implement (in software) the calculation 
of the angles $\psi_2$ and $\psi_{3 \pm}$, leading to the
evaluation of $({\rm Det}_2)^{1/2}$ and $({\rm Det}_{3 \pm})^{1/3}$,
where ${\rm Det}_2$ and ${\rm Det}_{3 \pm}$ are defined
in eqs.~(\ref{Det2}) and (\ref{Det3pm}). The running of these
quantities in their own complex planes, when $u$ ($\equiv |Q^2|/\Lambda^2$)
decreases from infinity towards zero, at fixed $\phi$ 
(we recall: $Q^2 = |Q^2| e^{i \phi}$, $-\pi < \phi \leq \pi$) is presented
in figs.~\ref{FigDet2}, \ref{FigDet3m}, \ref{FigDet3p}, respectively 
(in subsec.~\ref{subsec:num5l}). We note that the softwares, such as
Mathematica \cite{Math8}, assign to the complex numbers the arguments 
(angles) in the interval $]-\pi, \pi]$, i.e., they are not capable of 
keeping track of the correct arguments once those arguments go outside 
this interval. Therefore, in order to implement in practice the
unambiguous calculation of $\psi_2$ and $\psi_{3 \pm}$, we need to lift the
ambiguity $\psi_j \leftrightarrow \psi_j + 2 \pi k$ ($j=2, 3-, 3+$; $k$ is integer))  
obtained by the numerical evaluations of these arguments. 
Below we show how this ambiguity is lifted in practice in the 
case of subsec.~\ref{subsec:num5l},
i.e., the effective five-loop QCD case in ${\overline {\rm MS}}$ scheme
and with $n_f=3$.

\begin{enumerate}

\item
For ${\rm Det}_2(u, \phi) \equiv |{\rm Det}_2(u, \phi)| \exp(i \psi_2)$:
\begin{itemize}
\item
If $\phi_{\rm thr} \leq \phi \leq \pi$ (fig.~\ref{FigDet2}(a)), 
we can check that ${\rm Im} {\rm Det}_2(u_{\rm thr},\phi) < 0$.
Therefore: 
(a) when ${\rm Im} {\rm Det}_2 \geq 0$ and $u> u_{\rm thr}$, 
we have $0 \leq \psi_2  \leq \pi$; 
(b) when  ${\rm Im} {\rm Det}_2 \geq 0$ and $u <u_{\rm thr}$,
we have $2 \pi \leq \psi_2 \leq 3 \pi$; 
(c) when ${\rm Im} {\rm Det}_2 < 0$, we have
$\pi < \psi_2 < 2 \pi$. 
\item
If ($0 \leq$) $\phi < \phi_{\rm thr}$ (fig.~\ref{FigDet2}(b)), 
we have $-\pi/2 < \psi_2 < \pi$.
\end{itemize}
\item
For ${\rm Det}_{3-}(u, \phi) \equiv |{\rm Det}_{3-}(u, \phi)| \exp(i \psi_{3-})$:
\begin{itemize}
\item
If $\phi_{\rm thr} \leq \phi \leq \pi$ (fig.~\ref{FigDet3m}(a)), 
we can check that for $u_0 \equiv u_{\rm thr} + 0.001$ we have 
${\rm Im} {\rm Det}_{3 -}(u_0,\phi) < 0$ and that it represents a point in
the left sector of the two sectors with ${\rm Im} {\rm Det}_{3 -} < 0$
(see fig.~\ref{FigDet3m}(a); further, for $\phi$ sufficiently
large, there is only one sector with  ${\rm Im} {\rm Det}_{3 -} < 0$).
Therefore: 
(a) when ${\rm Im} {\rm Det}_{3 -} \geq 0$ and $u > u_0$, 
we have $0 \leq \psi_{3-} \leq \pi$; 
(b) when  ${\rm Im} {\rm Det}_{3 -} \geq 0$ and $u < u_0$, we have
$2 \pi \leq \psi_{3-} \leq 3 \pi$;
(c) when  ${\rm Im} {\rm Det}_{3 -} < 0$, we have $\pi < \psi_{3-} < 2 \pi$.
\item
If ($0 \leq$) $\phi <  \phi_{\rm thr}$ (fig.~\ref{FigDet3m}(b)),
we can check that the transition of ${\rm Det}_{3 -}$
from the 4th to the 5th quadrant
takes place at u between $u_4 \equiv u_{\rm thr}+0.0014$ and 
$u_5 \equiv u_{\rm thr}-0.0017$  ($u_5 < u_{4 \to 5} < u_4$); and that
for $u = u_7 \equiv 0.174$ the quantity ${\rm Det}_{3 -}$ is in the
7th quadrant. Therefore: 
(a) when ${\rm Im} {\rm Det}_{3 -} \geq 0$ and $u > u_4$, 
we have $0 \leq  \psi_{3-} \leq \pi$; 
(b) when  ${\rm Im} {\rm Det}_{3 -} \geq 0$ and $u_7 < u < u_4$, 
we have $2 \pi \leq  \psi_{3-} \leq 3 \pi$;
(b) when  ${\rm Im} {\rm Det}_{3 -} \geq 0$ and $u < u_7$, 
we have $4 \pi \leq  \psi_{3-} \leq 5 \pi$;
(d) when ${\rm Im} {\rm Det}_{3 -} < 0$ and $u > u_5$, 
we have $\pi <  \psi_{3-} < 2 \pi$;
(e) when ${\rm Im} {\rm Det}_{3 -} < 0$ and $u < u_5$, 
we have $3 \pi <  \psi_{3-} < 4 \pi$;
\end{itemize}
\item
For ${\rm Det}_{3+}(u, \phi) \equiv |{\rm Det}_{3+}(u, \phi)| \exp(i \psi_{3+})$:
\begin{itemize}
\item
If $\phi_{\rm thr} \leq \phi \leq \pi$ (fig.~\ref{FigDet3p}(a)), 
we can check that at $u_{\rm low}=0.174$ we have ${\rm Im}{\rm Det}_{3+} < 0$.
Therefore,
(a) when ${\rm Im} {\rm Det}_{3 +} \geq 0$ and $u > u_{\rm low}$, 
we have $0 \leq  \psi_{3+} \leq \pi$; 
(b) when ${\rm Im} {\rm Det}_{3 +} \geq 0$ and $u < u_{\rm low}$, 
we have $2 \pi \leq  \psi_{3+} \leq 3 \pi$; 
(c) when ${\rm Im} {\rm Det}_{3 +} < 0$,
we have $\pi <  \psi_{3+} < 2 \pi$.
\item
If ($0 \leq$) $\phi <  \phi_{\rm thr}$ (fig.~\ref{FigDet3p}(b)),
we have $-\pi/2 < \psi_{3+} < \pi$.
\end{itemize}
\end{enumerate}

When the number of effective quark flavors $n_f$ changes, 
or the renormalization scheme changes away from
${\overline {\rm MS}}$, the above rules 
in general do not get modified qualitatively, but quantitatively the
values $\phi_{\rm thr}$, $u_{\rm thr}$, $u_4$, $u_5$, etc. do change.

\end{document}